\documentclass{article}

\usepackage{microtype}
\usepackage{graphicx}
\usepackage{subcaption}
\usepackage{booktabs} 

\usepackage{hyperref}
\usepackage{xcolor}

\usepackage[accepted]{icml2026}

\usepackage{amsmath}
\usepackage{amssymb}
\usepackage{mathtools}
\usepackage{amsthm}

\usepackage[capitalize,noabbrev]{cleveref}

\theoremstyle{plain}

\theoremstyle{definition}

\theoremstyle{remark}

\usepackage[textsize=tiny]{todonotes}

\icmltitlerunning{Efficient Code Analysis via Graph Representation Learning-Guided Large Language Models}

\usepackage{tcolorbox}
\usepackage{multirow}
\usepackage[table]{xcolor}
\definecolor{codebg}{rgb}{0.95,0.95,0.95}
\definecolor{keyword}{rgb}{0.13,0.13,0.81}
\definecolor{string}{rgb}{0.81,0.13,0.13}
\definecolor{func}{rgb}{0.81,0.13,0.81}
\definecolor{DarkRed}{rgb}{0.55,0.0,0.0}

\begin{document}

\twocolumn[
  \icmltitle{Efficient Code Analysis via Graph Representation Learning-Guided Large Language Models}
  


  \icmlsetsymbol{equal}{*}
    
  \begin{icmlauthorlist}
    \icmlauthor{Hang Gao}{equal,yyy,sch}
    \icmlauthor{Tao Peng}{equal,comp,sch}
    \icmlauthor{Baoquan Cui}{yyy}
    \icmlauthor{Hong Huang}{yyy,sch}
    \icmlauthor{Fengge Wu}{comp,sch}
    \icmlauthor{Junsuo Zhao}{comp,sch}
    \icmlauthor{Jian Zhang}{yyy,sch}
  \end{icmlauthorlist}

  \icmlaffiliation{yyy}{Key Laboratory of System Software, Institute of Software, Chinese Academy of Sciences, Beijing, China}
  \icmlaffiliation{comp}{Institute of Software, Chinese Academy of Sciences, Beijing, China}
  \icmlaffiliation{sch}{University of Chinese Academy of Sciences, Beijing, China}

  \icmlcorrespondingauthor{Fengge Wu}{fengge@iscas.ac.cn}
  \icmlcorrespondingauthor{Baoquan Cui}{cuibq@ios.ac.cn}
  \icmlkeywords{Machine Learning, ICML}

  \vskip 0.3in
]



\printAffiliationsAndNotice{\icmlEqualContribution}


\begin{abstract}
Large Language Models (LLMs) have significantly advanced code analysis tasks, yet they struggle to detect malicious behaviors fragmented across files, whose intricate dependencies easily get lost in the vast amount of benign code. We therefore propose a graph-centric attention acquisition pipeline that enhances LLMs' ability to localize malicious behavior. The approach parses a project into a code graph, uses an LLM to encode nodes with semantic and structural signals, and trains a Graph Neural Network (GNN) under sparse supervision. The GNN performs an initial detection, and by interpreting these predictions, identifies key code sections that are most likely to contain malicious behavior. These influential regions are then used to guide the LLM's attention for in-depth analysis. This strategy significantly reduces interference from irrelevant context while maintaining low annotation costs. Extensive experiments show that the method consistently outperforms existing approaches on multiple public and custom datasets, highlighting its potential for practical deployment in software security scenarios. Codes can be found in https://github.com/Epiphaniespt/GMLLM.git.

\end{abstract}

\section{Introduction}

LLMs excel at handling code-related tasks \citep{nam2024using,haroon2025accurately}, including code understanding \citep{li2024mutation}, generation \citep{fan2023large}, and logic optimization \citep{ishibashi2024self,wei2024extending}. Current LLMs can generate efficient and highly readable code based on requirements, thanks to their deep understanding of a wide range of programming languages and syntax. Through natural language processing techniques, LLMs can also effectively identify syntax errors and potential vulnerabilities in code snippets and provide developers with suggestions for fixes, thereby improving code quality \citep{huynh2025detecting}. However, in the field of code security—particularly in malicious code detection—LLMs face limitations, as they are not yet capable of analyzing entire code projects with the same depth and comprehensiveness as security experts or professional tools, especially when it comes to large-scale projects \citep{wang2025malpacdetector,akuthota2023vulnerability,xue2024poster}.

\begin{figure}[t] 
  \centering
  \begin{minipage}{0.35\textwidth} 
    \centering
    \includegraphics[width=\linewidth]{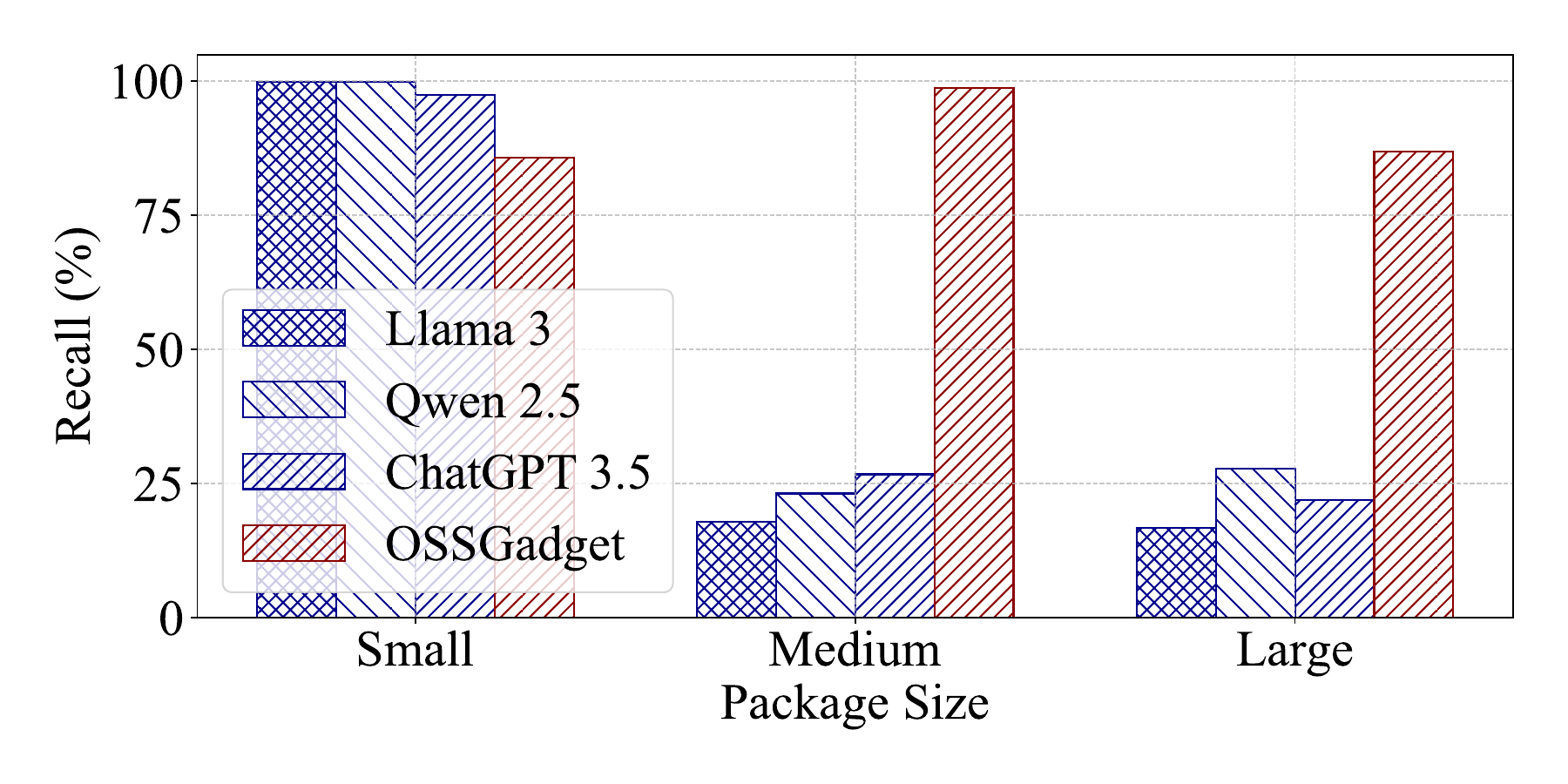}
    \subcaption{Recall}
  \end{minipage}
  \hfill 
  \begin{minipage}{0.35\textwidth}
    \centering
    \includegraphics[width=\linewidth]{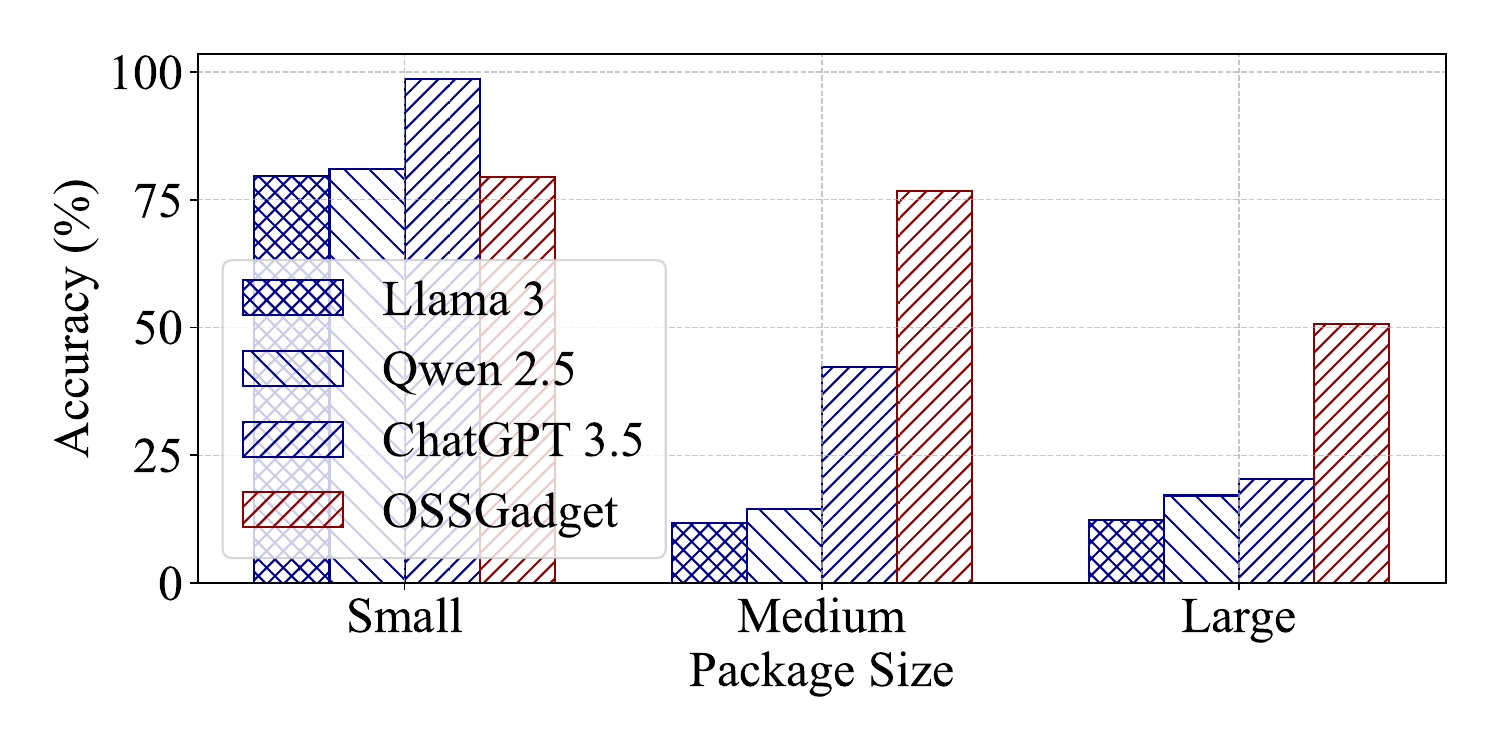}
    \subcaption{Accuracy}
  \end{minipage}
  
  \caption{Comparison between LLMs directly applied to malicious code detection and a commonly used malicious code detection tool OSSGadget \citep{microsoft_oss_detect_backdoor}. The blue bars represent the performance of the LLMs, while the red bar represents the performance of the tool. Large, Medium, Small refer to the scale of packages.}
  \label{fig:example}
\end{figure}

The above argument can be practically validated. Specifically, we conduct tests on some collected Python packages. The experimental results are shown in Figure \ref{fig:example}. As seen, the overall analysis performances of the LLM baselines are still far from those of professional analysis tools, especially for large packages. Another obvious observation is that, as the size of the package increases, the performance of LLM in detection decreases. We believe this is primarily due to the inherent characteristics of LLMs. When dealing with large amounts of code, LLMs' attention mechanisms often struggle to accurately pinpoint problematic snippets, instead wasting attention on large portions of benign code \citep{DBLP:conf/acl/HuHWLHLCXS25}. This explains why, in the experimental results shown in the figure, the larger the program, the worse the performance of the LLM analysis. Additionally, when the internal relationships within the code are complex and involve multiple interdependencies, LLMs face even greater difficulty. They struggle to identify malicious behaviors that are present in call relationships, especially when the relevant code snippets are not sequentially close. This also hinders their ability to properly attend to and jointly analyze these distant, yet connected, pieces of code. 

Furthermore, when LLMs are used to analyze a large number of code projects, they typically require significantly more computational resources compared to conventional methods. Unlike typical tools that operate based on specific rules or smaller, specialized neural networks, LLMs rely on their large-scale models and the vast amount of knowledge they contain to identify malicious code. This means that for each code project, especially at a large scale, resource consumption—such as memory and processing time—can increase rapidly. As a result, LLMs are less efficient in scenarios that require large-scale code analysis.

We aim to address these issues and enhance the capabilities of LLMs in malicious code detection, striving to develop a solution that surpasses professional tools in performance while being cost-effective, thereby highlighting the practical value of our research. To this end, our research focuses on Python. As an interpreted language, its nature, which is closer to that of a natural language, is inherently compatible with the sequence-processing mechanisms of LLMs \citep{wang2021codet5,DBLP:conf/iclr/NijkampPHTWZSX23}, making it an ideal subject for validating our approach. Additionally, Python code and technical documentation constitute some of the most abundant sources of code training data for LLMs, which leads to enhanced analysis capabilities \citep{DBLP:journals/corr/abs-2107-03374,DBLP:journals/corr/abs-2308-12950}. Therefore, towards Python code, we propose a \textit{\textbf{G}raph Representation Learning-Guided \textbf{M}alicious code detection framework for \textbf{LLM}}, abbreviated as GMLLM. GMLLM utilizes a graph representation learning approach to identify critical sections within projects that require closer inspection, which serves as attentions that allow the LLM to focus on these key areas, reducing redundant information and improving detection performance.

GMLLM first constructs graph representations of the input Python packages, using an LLM to extract feature vectors for each node. It then trains a GNN model with minimally labeled data that indicates whether a package contains malicious code. This trained model provides an initial, albeit potentially inaccurate, assessment of new packages. We then introduce an explanation paradigm to identify and measure the influence of input graph nodes on the GNN's classification. If a node has a significant influence on the GNN's decision to classify a sample as malicious, this suggests that the node itself possesses features characteristic of malicious code. We then filter these influential nodes and use an LLM for a more detailed analysis. This approach both fully leverages the LLM's code understanding capabilities and avoids the performance issues associated with analyzing entire packages. Furthermore, the method requires only binary labels (i.e., malicious or not) for training, which allows it to make full use of abundant relevant data. We compare this approach with multiple LLM baselines and other malicious code detection methods and demonstrate its superiority.

 Our contributions are as follows:

\begin{itemize}
\item We propose a novel paradigm to address the challenges LLMs face in detecting malicious code, thereby fostering the broader application of LLMs in software security-related domains.

\item Focusing on Python programs, we design the GMLLM framework to implement the proposed paradigm, which significantly enhances the malicious code detection capabilities of LLMs.

\item We validate our approach using a proposed comprehensive Python malicious code dataset and multiple public datasets, demonstrating the effectiveness of our method. 
\end{itemize}

\section{Related Works}

\subsection{Malicious Code Detection}
Malicious code detection is vital for information security, with conventional methods including metadata-based, rule-based, and learning-based approaches. Metadata-based methods, such as package name analysis \citep{DBLP:journals/corr/abs-2108-09576}, are easily bypassed. Rule-based methods, like Yara \citep{virustotal2023yara} and Bandit \citep{bandit2023}, are more precise but require expert-designed rules, making them resource-intensive \citep{DBLP:conf/ndss/DuanAKESL21}. Learning-based methods using deep learning have achieved high accuracy, extracting features such as API calls and opcodes \citep{DBLP:journals/compsec/YadavMRVP22, 2024AMDDLmodel, DBLP:journals/iet-ifs/WuSWZS23}, but face challenges with adversarial attacks \citep{DBLP:journals/iotj/YumlembamIJY23, DBLP:journals/ett/AminSSAKA22}. In the meantime, within open-source ecosystems like PyPI, malicious code attacks still cause severe disruptions and losses \citep{okafor2022sok}. While machine learning models have achieved near-perfect detection performance \citep{dambra2023decoding,DBLP:conf/kbse/LiangLWLW23,DBLP:conf/kbse/0001GCB0024}, challenges remain regarding methodology, dataset design, and false positive tolerance \citep{arp2022dos, DBLP:conf/icse/VuNM23}.

\subsection{LLM for Malicious Code Detection}
There is a growing trend toward leveraging LLMs for the detection of malicious code \citep{DBLP:conf/icse/ZahanBLAW25}. \citet{fang2024llm} find that LLMs perform well on simple code but struggle with complex or obfuscated code. \citet{yu2024maltracker} enhances detection accuracy by combining LLMs with static analysis. \citet{akinsowon2024leveraging} show LLMs improve detection rates for malicious code challenging for static analysis. LLMs also address malicious code sample scarcity by generating diverse datasets \citep{yu2024maltracker}, but their generative capabilities pose security risks. \citet{khan2024exploring} propose an integrated framework combining LLMs with traditional techniques to effectively address malicious code detection. Despite the advances, LLMs still face difficulties with large-scale code projects. Our work aims to address this issue.

\section{Method}
\begin{figure*}[htbp]
    \centering
    \includegraphics[width=0.9\textwidth]{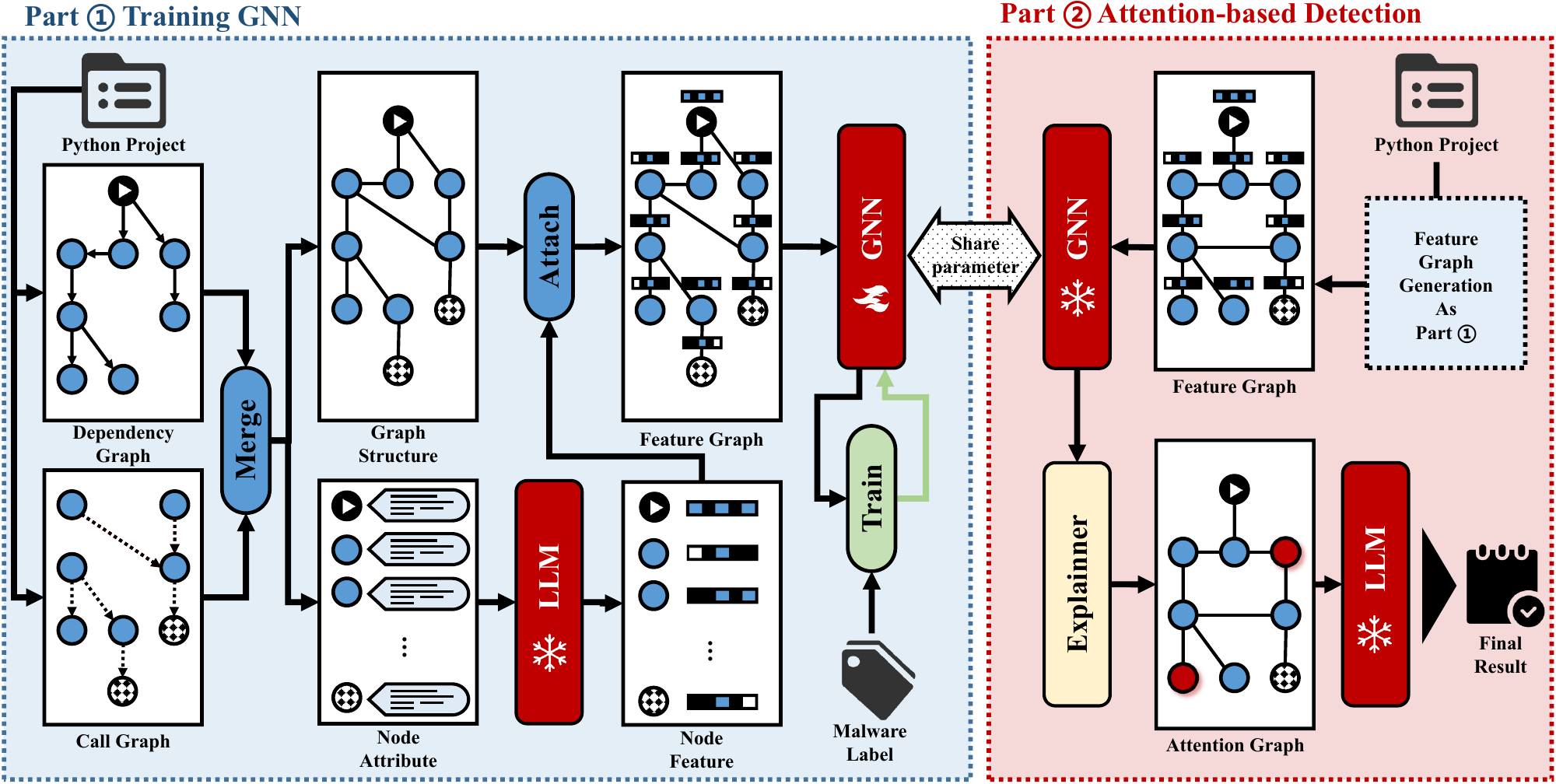}
    \caption{Framework of the proposed GMLLM.}
    \label{fig:framework}
\end{figure*}

The framework of our proposed GMLLM is shown in Figure \ref{fig:framework} and can be divided into two parts. Part one involves training a GNN using a dataset of malicious code with simple annotations. Part two interprets the trained GNN on new samples to obtain attention, which then guides the LLM to perform malicious code detection. Below, we will introduce these two parts separately.

\subsection{Training GNN}

To facilitate subsequent processing using GNNs, we represent the source code of each Python project as a graph \( G^{\text{code}} = \{\mathcal{V}^{\text{code}}, \mathcal{E}^{\text{code}}\} \), where \( \mathcal{V}^{\text{code}} \) is the set of nodes and \( \mathcal{E}^{\text{code}} \) is the set of edges.

The node set \( \mathcal{V}^{\text{code}} \) is derived from the Abstract Syntax Tree (AST) of the Python project. We traverse all the ``.py'' files and parse the source code into AST objects. The nodes in \( \mathcal{V}^{\text{code}} \) correspond to key entities in the Python code, such as classes, functions, and modules. Each node is annotated with the source code corresponding to the respective element. The edge set \( \mathcal{E}^{\text{code}} \) includes dependency edges and call relationship edges. Please refer to \textbf{Appendix} \ref{apx:GCD} for detailed implementations.

Next, we extract the features of each node \( v \) in \( \mathcal{V}^{\text{code}} \) and represent them as a vector \( h_v \). The feature extraction process involves defining a set of sensitive behavior rules \( \mathcal{S} \), which are used to capture the sensitive behavior features of the code corresponding to each node $v$. Such a rule-matching approach is commonly used in Python malicious code detection. What sets this approach apart is that we have employed LLM for automated rule generation and design the corresponding rule form. 

The set \( \mathcal{S} \) can be divided into two parts: common sensitive behavior rules \( \mathcal{S}^{\text{comm}} \), and data-derived sensitive behavior rules \( \mathcal{S}^{\text{data}} \). \( \mathcal{S}^{\text{comm}} \) is generated as follows:
\[
\mathcal{S}^{\text{comm}} = \text{LLM}(\rho^{\text{comm}}),
\]
where \( \text{LLM}(\cdot) \) represents the LLM model, and \( \rho^{\text{comm}} \) is the prompt used. \( \rho^{\text{comm}} \) instructs the LLM to summarize common sensitive behavior rules and output them as a list. 

The rules in \( \mathcal{S}^{\text{data}} \) are obtained by sampling 10\% of the training data and passing the corresponding code through the LLM to generate the data-related sensitive behavior rules:
\[
\mathcal{S}^{\text{data}} = \bigcup_{i=1}^{m} \text{LLM}(\rho^{\text{data}}, X_i)
\]
where \( \{X_i\}_{i=1}^{m} \) represents the set of Python files collected from the training data. Details concerning the implementations and prompts are offered in \textbf{Appendix} \ref{apx:rgd}. 

After obtaining \( \mathcal{S} \), a feature vector \( h_v \) of dimension \( |\mathcal{S}| \) is constructed for each node \( v \). This vector is a multi-hot vector where each position indicates whether a rule in \( \mathcal{S} \) is matched. This method automates the extraction of node-sensitive behavior features and is highly scalable. When new scenarios arise, the system only requires the addition of new environmental samples to automatically update the sensitive feature rules.

Based on the collected graph and node features, we proceed to train a GNN $g_{\theta}(\cdot)$. As discussed in the introduction, we formulate the task as a binary classification problem in order to minimize the need for labeled data. Following the aforementioned transformations, we obtain $n$ graph samples, denoted as $\{G_{i}^{\text{code}}\}_{i=1}^{n}$, where $n$ is the number of Python projects in the training set. These samples are then processed using a GNN. In our implementation, we employ a two-layer Graph Convolutional Network (GCN), though it is worth noting that any GNN model capable of handling graph-structured data could be applied to this task.

The training process follows a standard binary classification framework, with cross-entropy loss used for optimization. Formally, it can be expressed as:


\begin{equation}
\begin{split}
\mathcal{L}(\theta) = -\frac{1}{n} \sum_{i=1}^{n} \bigg[ & y_i \log \left( g_{\theta}(G_{i}^{\text{code}}) \right) \\
& + (1 - y_i) \log \left( 1 - g_{\theta}(G_{i}^{\text{code}}) \right) \bigg]
\end{split}
\end{equation}

\subsection{Attention Based Detection}
\label{sec:method_explainer}

We have developed a GNN model designed to detect potential malicious code based on sensitive behavioral features. This model is used to classify the target package, after which we conduct interpretability analysis on the classification results to identify the key nodes that lead the GNN to classify a package as malicious. In real-world projects, malicious code segments typically constitute only a small fraction of the overall codebase. Therefore, this approach can effectively pinpoint suspicious nodes, enabling an LLM to perform targeted analysis on these critical areas. In essence, we locate which specific nodes ``led'' the GNN to classify the target package as malicious, then focus on analyzing those nodes to verify whether they indeed represent malicious code and to understand their malicious behavior. The key advantage of this approach is that it provides the LLM with a focused attention mechanism for the entire project at relatively low training and inference costs, while avoiding the accuracy and resource consumption issues associated with having the LLM directly process the whole package.

Now we give a detailed presentation. For the case where the $j$-th sample under test, $G_{j}^{\text{code}}$, is identified as containing malicious code, we will specifically construct trainable edge mask $M_{j}^{\text{edge}}$ and a feature mask $M_{j}^{\text{feat}}$ on the target graph $G_{j}^{\text{code}}$. Let the adjacency matrix and feature matrix of $G_{j}^{\text{code}}$ be $A_{j}\in\{0,1\}^{|\mathcal{V}_{j}|\times |\mathcal{V}_{j}|}$ and $H_{j}\in\mathbb{R}^{|\mathcal{V}_{j}| \times |\mathcal{S}|}$. For the target graph $G_{j}^{\text{code}}$, we then set $M_{j}^{\text{edge}}\in\mathbb{R}^{|\mathcal{V}_{j}|\times |\mathcal{V}_{j}|}$ and $M_{j}^{\text{feat}}\in\mathbb{R}^{|\mathcal{V}_{j}|}$. Apply a sigmoid activation $\delta(\cdot)$ to $M_{j}^{\text{edge}}$ and symmetrize:
\begin{gather}
\dot{M}_{j}^{\text{edge}}=\delta(M_{j}^{\text{edge}}),\qquad \tilde M_{j}^{\text{edge}}=\frac{\dot{M}_{j}^{\text{edge}}+(M_{j}^{\text{edge}})^\top}{2},
\end{gather}
then remove the diagonal and perform an element-wise product with the original graph to obtain the \emph{masked adjacency}:
\begin{gather}
\hat A_{j}=\bigl(A_{j}\odot \tilde M_{j}^{\text{edge}}\bigr)\odot (\mathbf{1}-I),
\end{gather}
where $\odot$ denotes the Hadamard product. On the feature side, use $\tilde M^{\text{feat}}=\sigma(M^{\text{feat}})$ and broadcast:
\begin{gather}
H_{j}' = \text{diag}(\tilde M_{j}^{\text{feat}})H_{j}.
\end{gather}
Here, $\text{diag}(\tilde M_{j}^{\text{feat}})$ denotes a diagonal matrix whose diagonal entries are the elements of $\tilde M_{j}^{\text{feat}}$. Next, we feed $(H_{j}',\hat A_{j})$ into the trained GNN $g_\theta(\cdot)$ to obtain the predicted probabilities $p_{j}$ for sample $G_{j}^{\text{code}}$. $p_{j,(0)}$ and $p_{j,(1)}$ denotes the predicted probability of $g_\theta(\cdot)$ on $G_{j}^{\text{code}}$ as benign and malicious respectively. Then, we update the trainable masks using the following loss: 

\begin{gather}
\mathcal{L}_{\text{pred}} = -\log \left( p_{j,(1)} + \epsilon \right),
\end{gather}
where minimizing $\mathcal{L}_{\text{pred}}$ encourages the mask to preserve the substructure that maximize $p_{j,(1)}$. Here $\epsilon$ is a small constant for numerical stability.

To encourage sparsity and determinacy, we add size regularizers on the edge and feature masks:
\begin{gather}
\mathcal{L}_{\text{size}}=\|\tilde M_{j}^{\text{edge}}\|_{1} +\|\tilde M_{j}^{\text{feat}}\|_{1},
\end{gather}
and entropy regularizer:



\begin{equation}
\begin{split}
\mathcal{L}_{\text{ent}} = \frac{1}{|\tilde M_{j}^{\text{edge}}|} & \sum_{m_{(k,l)} \in \tilde M_{j}^{\text{edge}}} \Big( -m_{(k,l)}\log m_{(k,l)} \\
& -(1-m_{(k,l)})\log (1-m_{(k,l)}) \Big).
\end{split}
\end{equation}

The overall objective is:
\begin{gather}
\mathcal{L}=\mathcal{L}_{\text{pred}}+ \lambda^{\text{size}}\mathcal{L}_{\text{size}}+\lambda^{\text{ent}}\mathcal{L}_{\text{ent}}.
\end{gather}
$\mathcal{L}$ is optimized through backpropagation, yielding $M_{j}^{\text{edge}}$ and $M_{j}^{\text{feat}}$ as the explanatory mask. $\lambda^{\text{size}}$ and $\lambda^{\text{ent}}$ are hyperparameters.
Note that $\mathcal{L}$ is computed and optimized individually for each sample, with the objective of finding the appropriate mask. The masks $\tilde M_{j}^{\text{edge}}$ and $\tilde M_{j}^{\text{feat}}$ effectively reflect the graph structures that lead the GNN to classify the sample as containing malicious code, as these nodes maximize $p_{j,(1)}$, the predicted probability of $g_\theta(\cdot)$ on $G_{j}^{\text{code}}$ being malicious. This is precisely the part that should be analyzed in detail using LLMs. Therefore, $\mathcal{L}$ is applied on a per-sample basis.

We set the value of the attention score for each element of $G_{j}^{\text{code}}$ as the corresponding mask value within $\tilde M_{j}^{\text{edge}}$. Formally, we have the node attention score $a_{v}$ as:
\begin{gather}
    a_{v} = (\tilde M_{j}^{\text{feat}})_{v},
\end{gather}
and edge attention score $a_{v,w}$ as:
\begin{gather}
    a_{v,w} = (\tilde M_{j}^{\text{edge}})_{v,w}.
\end{gather}

Based on the acquired attention score, we employ an LLM to perform more targeted and in-depth analysis. We set a threshold $\gamma$, and only the code graph structures with scores greater than $\gamma$ are fed into the LLM for malicious code detection. In practice, since the majority of the code is benign, this step can filter out a large amount of irrelevant code, significantly saving computational resources and enabling the LLM to process the problem more efficiently. We denote the extracted subgraph as $\text{Att}(\cdot)$, which needs to be processed by the LLM. $\text{Att}(\cdot)$ can be formulated as:

\begin{equation}
    \begin{split}
        \text{Att}(G^{\text{code}}_{j}) = \Big\{& \big\{v \in \mathcal{V}_{j}^{\text{code}}|a_{v}>\gamma^{\text{node}}\big\}, \\
    & \big\{(v,w)\in \mathcal{E}_{j}^{\text{code}}|a_{v,w} > \gamma^{\text{edge}}\big\}\Big\}.
    \label{eq:K-ablation}
    \end{split}
\end{equation}

In the implementation, in order to convert the graph-structured data into a natural language description that can be understood and processed by the LLM, we combine the information from both the AST and the source code. Based on the structure of $\text{Att}(G^{\text{code}}_{j})$, we provide segmented descriptions of the nodes, edges, and the content of the nodes.  A concrete example is given below:
\small
\begin{tcolorbox}[
    colback=codebg,     
    colframe=black,     
    boxrule=0.5pt,      
    arc=4pt,            
    title=Examples of $\text{Att}(G^{\text{code}}_{j})$, 
    fonttitle=\bfseries
]
\textbf{Nodes:}

15Cent-999.0.1.setup

15Cent-999.0.1.setup.CustomInstal1

......

\textbf{Edges:}

10Cent11-999.0.4.setup.CustomInstall →

10Cent11-999.0.4.setup.CustomInstall.run

10Cent11-999.0.4.setup →

10Cent11-999.0.4.setup.CustomInstall

......

\textbf{Node Attributes:}

Codes of 15Cent-999.0.1.setup are

\quad  "   import re, sys, pathlib

\quad   \space \space  indicators = r"""

......

\end{tcolorbox}

Based on $\text{Att}(G^{\text{code}}_{j})$, the overall process can be formally expressed as follows:
\begin{gather}
(\hat{Y},T) = \text{LLM}\Big(\rho^{\text{ana}},\text{Att}(G^{\text{code}}_{j})\Big),
\label{eq:yt}
\end{gather}
where $\hat{Y}$ denotes the predicted classification, $T$ denotes the detailed description of location and characteristics concerning the malicious codes and will be left blank if $G^{\text{code}}_{j}$ is classified as benign, and $\rho^{\text{ana}}$ denotes the utilized prompt. Details concerning $\rho^{\text{ana}}$ are given in \textbf{Appendix} \ref{apx:ana}.


Next, our method can output the final results using the obtained $\hat{Y}$ and $T$ from equation \ref{eq:yt}. Compared to other commonly used tools and methods, our approach not only detects the presence of malicious code but also provides specific description of its characteristics and locations.
\section{Experiments}

In this section, we conduct multiple experiments with our proposed GMLLM and other baselines to answer the following research questions (RQs):

\quad \textbf{RQ1:} Does GMLLM lead to a significant improvement in the detection of malicious code by LLMs?

\quad \textbf{RQ2:} Is the performance of GMLLM superior to that of conventional malicious code detection tools?

\quad \textbf{RQ3:} Can GMLLM provide a comprehensive analysis of malicious code?

\quad \textbf{RQ4:} Does GMLLM help in reducing the computational overhead of LLMs for malicious code detection?

\subsection{Settings}
We performed a series of experiments using multiple datasets to validate the effectiveness of our method. Specifically, we utilized three publicly available datasets: Backstabbers \citep{ohm2020backstabber}, Datadog \citep{guarddog2023malicious}, and Mal-OSS \citep{DBLP:conf/kbse/GuoXLHFL23}. Additionally, we created a larger-scale dataset, the Malicious Codes from PYPI (MalCP) dataset, by incorporating samples collected from PYPI. 
The MalCP dataset is partitioned into three subsets based on program size: Large, Medium, and Small. An additional All category is included to represent the entire set of program samples. The MalCP dataset also incorporates detailed descriptions of malicious behaviors into its benchmark to evaluate the performance of LLM-based methods.
In addition, MalCP covers a diverse set of attack tactics, with three major categories (Execution, Impact, Credential Access/Collection) and a long tail of others, so that no single tactic dominates the dataset.
Detailed information about the MalCP dataset can be found in \textbf{Appendix} \ref{apx:dataset}.

For each benchmark dataset, we use an 80/20 train/test split at the package level. The split is performed independently for each dataset. For GMLLM, the GNN and data-derived sensitive behavior rules are constructed only from the training split, while all methods are evaluated on the same held-out test split. Additional details on split construction and preprocessing are provided in \textbf{Appendix} \ref{app:data-split}. 

For the comparison with the baselines, we evaluated the performance of our method against various LLMs and malicious code detection methods and tools. Specifically, we selected multiple different LLMs for direct application in malicious code detection, including Qwen 2.5 \citep{DBLP:journals/corr/abs-2412-15115}, Llama 2 \citep{DBLP:journals/corr/abs-2307-09288}, Llama 3 \citep{DBLP:journals/corr/abs-2407-21783}, ChatGPT 3.5 \citep{DBLP:journals/corr/abs-2303-10420}, and ChatGPT 4o \citep{DBLP:journals/corr/abs-2410-21276}. Additionally, we chose three rule-based common malicious code detection tools: Bandit4Mal \citep{vu2020fork}, OSSGadget \citep{microsoft_oss_detect_backdoor}, and Virustotal \citep{virustotal}, as well as three language model- and neural network-based malicious code detection tools: MPHunter \citep{DBLP:conf/kbse/LiangLWLW23}, Ea4mp \citep{DBLP:conf/kbse/0001GCB0024} and MalGuard \citep{malguard2025Gao}. To ensure statistical robustness, all experimental results are reported as the average across five independent trials.
We implemented our method using two base LLMs, Llama 3 and ChatGPT 4o, and compared their performance with the baselines. 

The experiments are organized as follows. RQ1--RQ2 evaluate detection performance on public benchmarks and MalCP (Tables~\ref{tab:multi datasets},~\ref{tab:our dataset}). RQ3 evaluates behavior-level explanation quality (Table~\ref{tab:description}), and RQ4 measures token usage and runtime (Table~\ref{tab:cost}). We further include ablations to isolate the effect of rules and the budgeted subgraph extraction.

\begin{table*}[htbp]
  \centering
  \tiny
  \caption{Performance on Backstabbers, Datadog, Mal-OSS Datasets. \textbf{Bold} indicates the best performance, while \underline{underline} indicates the second-best performance.}
  \label{tab:multi datasets}
  \setlength{\tabcolsep}{7.5pt} 
  \renewcommand{\arraystretch}{1.0}
  \begin{tabular}{lccccccccc}
    \toprule
    \multirow{2}{*}{Model} & \multicolumn{3}{c}{Backstabbers} & \multicolumn{3}{c}{Datadog} & \multicolumn{3}{c}{Mal-OSS}  \\
    \cmidrule(lr){2-4} \cmidrule(lr){5-7} \cmidrule(lr){8-10}
    & {Recall} & {Precision} & {ACC} & {Recall} & {Precision} & {ACC} & {Recall} & {Precision} & {ACC}   \\
    \midrule    
    ChatGPT 3.5      & 28.45 & 37.19 & 44.66 & 54.30 & 48.32 & 56.30 & 73.34 & 60.14 & 65.29 \\
    ChatGPT 4o       & 92.89 & 90.65 & 92.28 & 80.07 & 84.73 & 85.53 & 89.69 & \underline{90.22} & \underline{90.76} \\
    Llama 2          & 3.48  & 3.20  & 6.65  & 15.12 & 11.03 & 12.88 & 11.78 & 10.14 & 11.14  \\
    Llama 3          & 24.96 & 22.45 & 25.39 & 43.99 & 30.33 & 33.86 & 72.31 & 45.17 & 46.74 \\
    Qwen 2.5         & 28.30 & 25.10 & 27.74 & 49.83 & 33.49 & 37.19 & 78.50 & 47.46 & 50.00  \\
    Bandit4Mal       & 12.92 & 29.67 & 45.53 & 62.89 & 66.06 & 70.77 & 23.56 & 45.33 & 51.63 \\
    OSSGadget        & \textbf{96.81} & 64.51 & 73.88 & 86.60 & 57.40 & 67.29 & 86.01 & 59.47 & 66.51 \\
    Virustotal       & 87.52 & 84.81 & 86.97 & 82.13 & 82.70 & 85.24 & 46.69 & 76.39 & 68.75 \\
    MPHunter         & 93.12 & 89.81 & 91.08 & 93.38 & \underline{90.17} & \underline{94.67} & 86.14 & 83.06 & 88.50 \\ 
    Ea4mp            & 94.58 & \underline{91.13} & \underline{93.17} & 82.13 & 82.70 & 85.24 & 85.19 & 83.72 & 84.99 \\
    MalGuard         & 53.85 & 67.33 & 66.55 & 79.38 & 73.10 & 79.02 & 87.48 & 76.06 & 81.52 \\
    \midrule
    GMLLM (Llama 3 based)  & \underline{96.52} & 74.30 & 82.94 & \textbf{97.94} & 73.83 & 84.52 & \textbf{98.82} & 68.89 & 78.87  \\
    GMLLM (ChatGPT 4o based)   & 93.03 & \textbf{94.54} & \textbf{94.29} & \underline{97.25} & \textbf{95.61} & \textbf{96.96} & \underline{96.76} & \textbf{95.36} & \textbf{96.33}  \\
    \bottomrule
  \end{tabular}
  \vskip -0.1in
\end{table*}

\begin{table*}[htbp]
  \centering
  \tiny
  \caption{Performance on our proposed MalCP dataset. \textbf{Bold} indicates the best performance, while \underline{underline} indicates the second-best performance.}
  \label{tab:our dataset}
  \setlength{\tabcolsep}{3.5pt} 
    \renewcommand{\arraystretch}{1.0}
  \begin{tabular}{lccccccccccccc}
    \toprule
    \multirow{2}{*}{Model} & \multicolumn{3}{c}{Large} & \multicolumn{3}{c}{Medium} & \multicolumn{3}{c}{Small} & \multicolumn{3}{c}{All} & \multicolumn{1}{c}{Benign} \\
    \cmidrule(lr){2-4} \cmidrule(lr){5-7} \cmidrule(lr){8-10} \cmidrule(lr){11-13} \cmidrule(lr){14-14}
    & {recall} & {precision} & {acc}  & {recall} & {precision} & {acc}  & {recall} & {precision} & {acc}& {recall} & {precision} & {acc}  & {specificity} \\
    \midrule
    ChatGPT 3.5     & 21.91 & 15.86 & 20.43 & 26.70 & 34.97 & 42.25 & 97.38 & \textbf{99.66} & \underline{98.59} & 51.36 & 50.65 & 55.18 & 58.35 \\
    ChatGPT 4o      & 67.25 & 76.17 & 77.98 & \underline{98.30} & \underline{93.53} & \underline{96.01} & 97.38 & 94.14 & 95.84 & 89.33 & 89.49 & 90.39 & 91.27 \\
    Llama 2         & 7.59  & 5.22  & 5.68  & 2.55  & 2.27  & 2.56  & 16.07 & 16.23 & 20.09 & 8.92  & 7.64  & 9.64  & 10.24 \\
    Llama 3         & 16.70 & 11.31 & 12.26 & 17.86 & 14.46 & 11.82 & \textbf{99.84} & 70.16 & 79.59 & 47.68 & 34.77 & 35.60 & 25.54 \\
    Qwen 2.5        & 27.77 & 17.58 & 17.14 & 23.13 & 18.01 & 14.46 & \textbf{99.84} & 71.65 & 81.00 & 52.62 & 37.42 & 38.50 & 26.74 \\
    Bandit4Mal      & 64.21 & 51.48 & 60.57 & 9.35  & 27.64 & 45.93 & 13.28 & 51.92 & 52.59 & 26.04 & 46.45 & 52.77 & 75.01 \\
    OSSGadget       & 86.77 & 44.74 & 50.71 & \textbf{98.64} & 67.21 & 76.76 & 85.74 & 74.93 & 79.43 & 90.60 & 61.22 & 69.66 & 52.23 \\
    Virustotal      & 83.51 & 85.37 & 87.39 & 87.93 & 84.89 & 86.98 & 42.13 & 71.99 & 64.44 & 69.86 & 81.79 & 79.24 & 87.05 \\
    MPHunter        & 85.60 & 81.27 & 83.83 & 93.32 & 89.18 & 90.52 & 96.39 & 95.03 & 96.50 & 94.13 & 90.81 & \underline{93.56} & 91.57\\ 
    Ea4mp           & \underline{87.31} & \underline{86.23} & \underline{88.10} & 91.76 & 92.63 & 92.79 & 97.19 & 96.74 & 97.00 & 93.12 & \underline{92.17} & 91.03 & \underline{92.37}  \\
    MalGuard        & 68.11 & 66.11 & 72.65 & 45.07 & 65.43 & 63.02 & \underline{99.67} & 80.21 & 88.07 & 71.55 &72.47 & 74.73 & 77.37 \\
    \midrule
    GMLLM (Llama 3 based)   & \textbf{94.36} & 65.32 & 77.18 & \underline{98.30} & 73.44 & 82.51 & \underline{99.67} & 75.81 & 84.62 & \textbf{97.71} & 71.88 & 81.60 & 68.19 \\
    GMLLM (ChatGPT 4o based)    & 87.20 & \textbf{89.14} & \textbf{90.41} & 97.45 & \textbf{96.14} & \textbf{96.96} & 99.34 & \underline{98.38} & \textbf{98.90} & \underline{95.30} & \textbf{95.07} & \textbf{95.62} & \textbf{95.89} \\
    \bottomrule
  \end{tabular}
  \vskip -0.1in
\end{table*}

\begin{table*}[!htbp]
  \centering
  \tiny
  \caption{Model performances on describing the details concerning the malicious behaviors. \textbf{Bold} indicates the best performance, while \underline{underline} indicates the second-best performance. Baselines including Bandit4Mal, OSSGadget, Virustotal, MPHunter, and Ea4mp are excluded as they are incapable of such description.}
  \label{tab:description}
  \setlength{\tabcolsep}{3.5pt} 
  \renewcommand{\arraystretch}{1.0}
  \begin{tabular}{lcccccc}
    \toprule
    \multirow{2}{*}{Method}
    & {Threat} & {Execution} & {Evidence } & {Average} & {Quality Score} & {Factual} \\
    & {Generality $(\uparrow)$} & {Path Traceability $(\uparrow)$} & {Groundedness $(\uparrow)$} & {Quality Score} $(\uparrow)$& {Standard Deviation $(\downarrow)$} & {Alignment} $(\uparrow)$ \\
    \midrule
    \multicolumn{7}{c}{All} \\
    \midrule
    ChatGPT 3.5 & 2.013 & 1.500 & 1.681 & 1.731 & 1.699 & 1.523 \\
    ChatGPT 4o & \underline{3.427} & \textbf{2.764} & \underline{2.967} & \underline{3.053} & 1.166 & \underline{2.602} \\
    Llama 2 & 0.244 & 0.177 & 0.207 & 0.209 & \underline{0.714} & 0.149 \\
    Llama 3 & 1.731 & 1.256 & 1.564 & 1.517 & 1.605 & 1.325 \\
    Qwen 2.5 & 1.995 & 1.120 & 1.510 & 1.541 & 1.487 & 1.420 \\
    \midrule
    GMLLM (Llama 3 based) & 3.037 & 2.056 & 2.464 & 2.519 & \textbf{0.647} & 2.060 \\
    GMLLM (ChatGPT 4o based) & \textbf{3.687} & \underline{2.727} & \textbf{3.215} & \textbf{3.210} & 0.860 & \textbf{2.721} \\
    \midrule
    \multicolumn{7}{c}{Large} \\
    \midrule
    ChatGPT 3.5 & 0.842 & 0.633 & 0.705 & 0.727 & 1.385 & 0.640 \\
    ChatGPT 4o & 2.716 & \underline{2.167} & 2.336 & \underline{2.406} & 1.732 & \underline{1.978} \\
    Llama 2 & 0.156 & 0.119 & 0.137 & 0.137 & \textbf{0.524} & 0.111 \\
    Llama 3 & 0.584 & 0.432 & 0.527 & 0.514 & 1.165 & 0.453 \\
    Qwen 2.5 & 1.017 & 0.640 & 0.777 & 0.811 & 1.349 & 0.714 \\
    \midrule
    GMLLM (Llama 3 based) & \underline{2.857} & 1.928 & \underline{2.369} & 2.385 & \underline{0.824} & 1.915 \\
    GMLLM (ChatGPT 4o based) & \textbf{3.377} & \textbf{2.520} & \textbf{3.045} & \textbf{2.981} & 1.262 & \textbf{2.512} \\
    \midrule
    \multicolumn{7}{c}{Medium} \\
    \midrule
    ChatGPT 3.5 & 1.024 & 0.748 & 0.845 & 0.872 & 1.462 & 0.777 \\
    ChatGPT 4o & \underline{3.682} & \textbf{2.794} & \underline{2.889} & \underline{3.122} & 0.637 & \underline{2.643} \\
    Llama 2 & 0.073 & 0.051 & 0.063 & 0.062 & \textbf{0.397} & 0.039 \\
    Llama 3 & 0.713 & 0.527 & 0.641 & 0.627 & 1.289 & 0.536 \\
    Qwen 2.5 & 0.859 & 0.505 & 0.658 & 0.674 & 1.250 & 0.582 \\
    \midrule
    GMLLM (Llama 3 based) & 2.917 & 1.959 & 2.287 & 2.388 & \underline{0.585} & 1.879 \\
    GMLLM (ChatGPT 4o based) & \textbf{3.755} & \underline{2.770} & \textbf{3.155} & \textbf{3.226} & 0.672 & \textbf{2.661} \\
    \midrule
    \multicolumn{7}{c}{Small} \\
    \midrule
    ChatGPT 3.5 & 3.851 & \underline{2.879} & 3.225 & 3.318 & 0.597 & 2.908 \\
    ChatGPT 4o & \underline{3.852} & \textbf{3.310} & \textbf{3.628} & \textbf{3.577} & 0.733 & \textbf{3.034} \\
    Llama 2 & 0.474 & 0.343 & 0.398 & 0.405 & 0.984 & 0.285 \\
    Llama 3 & 3.579 & 2.580 & 3.238 & 3.132 & \underline{0.493} & 2.746 \\
    Qwen 2.5 & 3.828 & 2.075 & 2.885 & 2.930 & 0.688 & 2.762 \\
    \midrule
    GMLLM (Llama 3 based) & 3.289 & 2.246 & 2.707 & 2.747 & \textbf{0.463} & 2.283 \\
    GMLLM (ChatGPT 4o based) & \textbf{3.853} & 2.842 & \underline{3.399} & \underline{3.365} & 0.562 & \underline{2.935} \\
    \bottomrule
  \end{tabular}
  \vskip -0.1in
\end{table*}

\begin{table*}[htbp]
\centering
\tiny
\caption{
A comparison of the token counts used by LLM baselines in the MalCP dataset, categorized by program sample size (Large, Medium, and Small). The best performance is indicated in \textbf{bold}, while the second-best is \underline{underlined}.
}
\label{tab:cost}
\setlength{\tabcolsep}{4pt} 
\renewcommand{\arraystretch}{1.0} 
\begin{tabular}{lrrrrrrrrrrrr}
\toprule
& \multicolumn{4}{c}{Large} & \multicolumn{4}{c}{Medium} & \multicolumn{4}{c}{Small} \\
\cmidrule(lr){2-5} \cmidrule(lr){6-9} \cmidrule(lr){10-13}
& Mean & Std & Min & Max & Mean & Std & Min & Max & Mean & Std & Min & Max  \\
\midrule
Llama3 & 259089.13 & 907541.88 & 365 & 12810537 & 28578.12 & 16497.45 & 296 & 39274 & 698.47 & 188.77 & 257 & 2523 \\
ChatGPT 4o & 251001.56 & 868507.92 & 370 & 12875186 & 28866.54 & 17267.57 & 307 & 150823 & 700.48 & 187.67 & 269 & 2544 \\
ChatGPT 3.5 & 255755.00 & 882059.26 & 376 & 12813070 & 29466.69 & 17614.16 & 307 & 150331 & 713.57 & 189.33 & 269 & 2530 \\
Qwen 2.5 & 280463.08 & 972292.04 & 375 & 14226559 & 29371.04 & 16789.81 & 297 & 40199 & 706.27 & 197.05 & 261 & 2533 \\
Llama 2 & 333972.02 & 1207832.80 & 482 & 1954134 & 31678.23 & 17814.37 & 385 & 43165 & 852.12 & 249.55 & 332 & 3297 \\
\midrule
GMLLM (Llama 3 based) & \underline{644.30} & \textbf{408.97} & \underline{201} & \underline{2587} & \underline{455.53} & \underline{200.56} & \textbf{199} & \underline{1661} & \underline{299.80} & \underline{254.70} & \textbf{204} & \underline{1698} \\
GMLLM (ChatGPT 4o based)  & \textbf{640.06} & \underline{409.35} & \textbf{198} & \textbf{2418} & \textbf{451.31} & \textbf{199.03} & \underline{203} & \textbf{1557} & \textbf{292.65} & \textbf{240.82} & \underline{209} & \textbf{1625} \\
\bottomrule
\end{tabular}
\vskip -0.1in
\end{table*}

\begin{figure*}
  \centering
  \begin{minipage}{0.3\textwidth}
    \includegraphics[width=\linewidth]{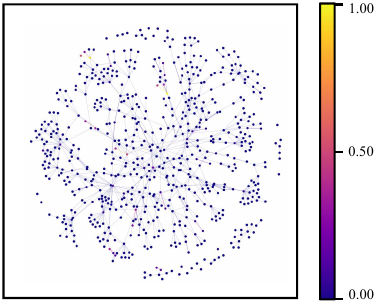}
    \subcaption{Large}
  \end{minipage}
  \begin{minipage}{0.3\textwidth}
    \includegraphics[width=\linewidth]{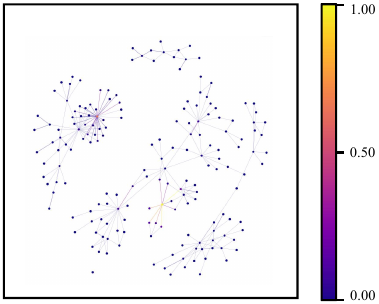}
    \subcaption{Medium} 
  \end{minipage}
  \begin{minipage}{0.3\textwidth}
    \includegraphics[width=\linewidth]{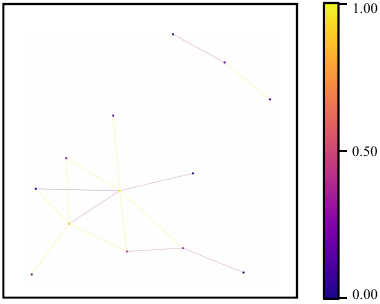}
    \subcaption{Small} 
  \end{minipage}
  \caption{A visualization of the code graph attention outputted with GMLLM when processing programs of different scales. In these graphs, the nodes and edges represent code elements and their corresponding relationships. The color intensity signifies the magnitude of attention, where colors closer to yellow indicate higher attention values. }
  \label{fig:vis} 
\end{figure*}

\subsection{Malicious Code Detection Capability Comparisons (RQ1, RQ2)}
We first compared the accuracy of malicious code detection. Tables \ref{tab:multi datasets} and \ref{tab:our dataset} present the detailed results. As observed, our method demonstrates superior performance on both public datasets and our own constructed dataset, outperforming baseline methods in the majority of scenarios and metrics. Notably, the improvement of our method over LLMs is significant, especially in the Precision metric, highlighting the ability of GMLLM to filter out redundant benign code interference, thereby validating the effectiveness of GMLLM. Furthermore, the comparison with other tools also indicates that, based on GMLLM, the LLM-based approach can effectively handle malicious code detection, particularly in the domain of Python code. 
This suggests that graph-guided prompting can close much of the gap between LLM-based approaches and specialized detectors, while retaining the ability to produce human-readable evidence for triage.
Interestingly, on small packages where baseline LLMs already achieve near-perfect recall, GMLLM maintains comparable performance, indicating that the graph-guided mechanism does not introduce unnecessary overhead when code complexity is low. This absence of regression validates the generality of our approach across varying project scales. 
To examine whether the effectiveness of GMLLM depends on the specific LLM backbones used in the main experiments, we further evaluate it with two additional strong LLM backbones, Qwen3.5-Flash and DeepSeek-V3.2, on a stratified subset of MalCP. The results, reported in \textbf{Appendix} \ref{apx:recent-llms}, show consistent improvements in both detection performance and token efficiency.

\subsection{Malicious Behavior Detailed Description Comparisons (RQ3)}

To analyze the performance of our algorithm, we compared the model-generated descriptions of malicious behavior against the ground-truth descriptions in the MalCP dataset. This comparison was limited to LLM-based methods, as other approaches are incapable of generating such detailed descriptions. For metrics, we used the following metrics: 1) Threat Generality, which measures the ability to generalize from specific functions or code elements to recognized cybersecurity tactics; 2) Execution Path Traceability, which assesses the clarity of the reconstructed execution flow; 3) Evidence Groundedness, which examines how closely a conclusion is supported by specific code elements; 4) Average Quality Score, the average score of the first three metrics; 5) Quality Score Standard Deviation, the standard deviation of the average evaluation quality; and 6) Factual Alignment, which evaluates the accuracy of identified and described malicious behaviors against a ground truth. Threat Generality, Execution Path Traceability, Evidence Correlation, and Factual Consistency were evaluated by an LLM, with the details discussed in \textbf{Appendix} \ref{apx:judge}.

The results, as shown in Table \ref{tab:description}, indicate that our method achieves superior performance in the vast majority of scenarios. This is particularly evident on large and medium-scale code, where foundational LLMs, when used alone, are more susceptible to interference.
To validate the reliability of using GPT-4o as an automatic evaluator for the explanation-quality metrics, we additionally performed a human evaluation study. Two security practitioners independently scored a subset of model explanations along the same four dimensions as in Table~\ref{tab:description}, and their
scores show consistent agreement with GPT-4o’s ratings (see \textbf{Appendix} \ref{apx:humanjudge} for detailed statistics). This suggests that GPT-4o is a reasonable proxy for expert judgment in our setting, while the core \emph{Factual Alignment} metric already relies purely on expert-curated labels.

\subsection{LLM Resource Consumption (RQ4)}
\label{sec:token}
We compared the resource consumption of GMLLM against other LLMs, using the number of tokens consumed as the primary metric. Calculated via model-specific tokenizers, this metric strictly accounts for system prompts and formatting overheads (see Appendix \ref{apx:token_methodology}). The results are presented in Table \ref{tab:cost}. 
As the data shows, GMLLM demonstrates the lowest resource usage in all scenarios, and the margin of improvement is significant. On \textsc{Large} packages, vanilla prompting exhibits highly unstable token counts, with heavy-tailed behavior (large standard deviation and extreme maxima). In contrast, GMLLM bounds the input length by design, leading to consistently low and predictable token costs across all size buckets and often reducing token usage by orders of magnitude on \textsc{Large}.
Beyond token usage, we also profile the cost of the explainer-based mask optimization used for subgraph extraction. On MalCP, this step takes about 10--11 seconds per package on average (see \textbf{Appendix}~\ref{apx:runtime}).

\subsection{Ablation Studies}
\label{sec:ablation}

\subsubsection{Effect of Rule Design and Model Components}

\begin{table}[htbp]
    \tiny
    \centering
    \caption{Ablation study of different rule configurations. \textbf{Bold} indicates the best performance.}
    \label{tab:ablation_results_Rule}
    \begin{tabular}{lcccc}
        \toprule
        Configuration & Recall & Precision & Accuracy & Benign Recall \\
        \midrule
        Structure-only (No $\mathcal{S}$) & 87.06 & 87.85 & 88.65 & 89.98 \\
        Human Rules ($\mathcal{S}^{\text{comm}}$) & 90.43 & 91.53 & 91.86 & 93.04 \\
        50 Rules ($\mathcal{S}^{\text{comm}} \cup 0.5\mathcal{S}^{\text{data}}$) & 92.78 & 93.28 & 93.69 & 94.44 \\
        Full GMLLM ($\mathcal{S}^{\text{comm}} \cup \mathcal{S}^{\text{data}}$) & \textbf{95.30} & \textbf{95.07} & \textbf{95.62} & \textbf{95.89} \\
        \bottomrule
    \end{tabular}
    \vskip -0.1in
\end{table}

To understand where the gains of GMLLM come from, we first ablate the rule set $\mathcal{S}$ and the model components on MalCP (GPT-4o backend). Table~\ref{tab:ablation_results_Rule} summarizes the effect of different rule configurations. ``Structure-only'' uses only structural graph features without any behavior rules. ``Human Rules'' ($\mathcal{S}^{\text{comm}}$) uses our manually designed common rules only. ``50\% Rules'' randomly keeps half of the rules in $\mathcal{S}^{\text{data}}$ at inference time, and ``Full GMLLM'' uses the complete rule set. In all cases, $\mathcal{S}^{\text{comm}}$ and $\mathcal{S}^{\text{data}}$ are fixed; we do not regenerate rules from different fractions of the dataset.

The results indicate that behavior rules provide substantial but not exclusive benefits. Without any behavior rules, the graph-only variant still achieves 88.65\% accuracy, showing that the structural representation itself is informative. Using only human-curated rules $\mathcal{S}^{\text{comm}}$ increases accuracy to 91.86\%, already exceeding the GPT-4o direct baseline on MalCP (90.39\%), which demonstrates that the framework does not rely solely on LLM-generated rules. Further enriching the rule set to $\mathcal{S}^{\text{comm}} \cup \mathcal{S}^{\text{data}}$ yields a monotonic improvement, with accuracy ultimately reaching 95.62\%.

We also compare a GNN-only classifier with the LLM-only and full GMLLM variants on MalCP. The GPT-4o direct baseline from Table~\ref{tab:our dataset} attains 90.39\% accuracy, the GNN-only model reaches 92.22\%, and the full GMLLM attains 95.62\%, corresponding to a 43.7\% relative reduction in error rate (7.78\% $\rightarrow$ 4.38\%). This shows that (i) the GNN with rule-augmented graphs already forms a strong detector, and (ii) the LLM provides complementary gains on top of the GNN, particularly on harder cases.

\begin{table}[htbp]
\tiny
\centering
\caption{Effect of top-$K$ edge budget on detection performance and token usage on MalCP (GPT-4o). Note that token cost is reported specifically for the \textbf{Large} dataset bucket.}
\label{tab:k_ablation}
\begin{tabular}{lccccc}
\toprule
$K$ (Max Edges) & Accuracy & Recall & Precision & Benign Recall & Token Cost \\
\midrule
10   & 92.97 & 94.18 & 90.75 & 91.96 & \textbf{435.45} \\
20   & \textbf{95.62} & \textbf{95.30} & \textbf{95.07} & \textbf{95.89} & 640.06 \\
30   & 95.19 & 95.45 & 94.03 & 94.96 & 810.33 \\
50   & 94.88 & 95.99 & 92.94 & 93.97 & 1131.33 \\
\bottomrule
\end{tabular}
\vskip -0.1in
\end{table}

\subsubsection{Sensitivity to the Subgraph Budget $K$}
\label{sec:k_ablation}

Rather than using a fixed attention threshold, we control the amount of code passed to the LLM with a budget $K$. In our implementation, this is achieved via a budgeted top-$K$ edge strategy: for each package, all call-graph edges are ranked by their explainer attention scores, and only the $K$ highest-scoring edges are retained to induce the subgraph. The same $K$ is shared across datasets.

We study the sensitivity of GMLLM to this budget on MalCP with GPT-4o by varying $K \in \{10, 20, 30, 50\}$ while keeping the GNN and explainer fixed. As shown in Table~\ref{tab:k_ablation}, increasing $K$ from 10 to 20 yields a clear performance gain (about +2.6 points in accuracy and +3.9 points in benign recall), suggesting that very small subgraphs may miss important context. For $K \ge 20$, however, detection metrics change only marginally, whereas the average token usage on Large packages grows almost linearly with $K$. We therefore fix $K=20$ in all main experiments as a reasonable operating point that balances detection performance and token cost.



\subsection{Deeper Insight}

To further interpret the scale-dependent performance pattern in Table~\ref{tab:our dataset}, we visualize the graph attention generated by GMLLM in Figure~\ref{fig:vis}. As illustrated, for the large-scale code, nearly all nodes are deep purple, which represents near-zero attention values. Only a small fraction of nodes and edges exhibit high attention, indicated by colors approaching yellow. This suggests that in larger-scale code projects, the proportion of code relevant to the malicious behavior is likely very small, which in turn explains the poor performance of LLMs on such samples. Our previous experiments corroborate this finding, underscoring the necessity of the additional attention mechanism in GMLLM. In contrast, on the small-scale dataset, the proportion of high-attention (yellow) nodes and edges increases significantly. 

This pattern also aligns with our observation that GMLLM does not regress on Small dataset. When code complexity is low, informative structures tend to be denser and more coherent, so graph guidance avoids over-filtering and preserves strong performance, supporting the generality of our approach.

\section{Conclusion}
We introduce GMLLM, a novel LLM-driven framework for malicious code detection designed to overcome a key limitation of LLMs: their susceptibility to noise from irrelevant code in large projects. Our approach leverages a lightweight GNN, trained on easily accessible data, which is interpreted during inference to produce attentions. These attentions direct the LLM to focus on potentially malicious code segments. As a result, GMLLM outperforms current state-of-the-art tools, rendering the direct use of LLMs for malicious code detection a practical possibility in real-world settings.


\section*{Impact Statement}

This paper presents work whose goal is to advance the field of Machine Learning for software supply chain security. The primary positive impact of our work is the enhancement of detection capabilities against malicious packages in open-source ecosystems (e.g., PyPI), potentially preventing security breaches and protecting developers.

However, we acknowledge potential negative societal consequences. First, false positives in our detection system could incorrectly flag legitimate packages, potentially harming the reputation of innocent developers. To mitigate this, we emphasize that our tool is designed to assist human experts rather than replace them. Second, there is a risk of dual-use: adversarial actors could potentially leverage the explainability features of our model to understand detection patterns and craft more evasive malware. We believe that the defensive benefits of transparency for security analysts outweigh these risks. Finally, the use of LLMs involves significant computational resources, contributing to energy consumption; we have attempted to mitigate this by using lightweight graph representations where possible.

\section*{Acknowledgments}
The authors would like to express their sincere gratitude to the anonymous reviewers for their insightful feedback and to the project team for their essential collaboration and support. This work was supported by the Beijing Natural Science Foundation under Grant No. 4264133, the Beijing Natural Science Foundation under Grant No. 4262072, the CAS Project for Young Scientists in Basic Research through Grant No. YSBR-040, and the Postdoctoral Fellowship Program of CPSF under Grant No. GZC20251642.


\bibliography{icml2026_conference}
\bibliographystyle{icml2026}

\nocite{langley00}

\newpage
\appendix
\onecolumn
\section{Usage of Large Language Model}
In our paper, we used LLMs to assist with polishing the writing, including correcting grammatical errors and making the sentences more consistent with academic English writing conventions.

\section{Extended Related Works}

\subsection{Malicious Code Detection Methods}
Malicious code detection is crucial for protecting information security, preventing data leaks, and defending against cyberattacks. It effectively prevents system damage and the loss of sensitive information. Conventional malicious code detection can be divided into three main categories: metadata-based malicious code detection, rule-based malicious code detection, and learning-based malicious code detection.

Metadata-based malicious code detection relies on discovering malicious code through information such as the package name and attributes. For example, Neupane et al. \citep{DBLP:conf/uss/NeupaneHWDC23} and Taylor et al. \citep{DBLP:conf/nss/TaylorVDCR20} use package names to identify malicious code. This approach, however, is relatively weak, as attackers can easily bypass it by mimicking the names of legitimate code packages \citep{DBLP:journals/corr/abs-2108-09576}. Existing work \citep{DBLP:conf/eurosp/VuPMPS20} attempt to optimize malicious code detection by calculating the edit distance between package names, but this may lead to a high false positive rate.

In contrast to the aforementioned methods, rule-based approaches are more detailed but require experts to manually design the rules, making them more resource-intensive. Specifically, there are existing rule-based detection tools for interpreted languages, including Yara \citep{virustotal2023yara}, Bandit \citep{bandit2023}, OSSGadget \citep{microsoft2023}, and OSSF \citep{ossf2023}. These tools are highly efficient. Furthermore, Duan et al. \citep{DBLP:conf/ndss/DuanAKESL21} and Huang et al. \citep{DBLP:conf/uss/0003WWSLCZHYS24} have combined static and dynamic analysis with related rules for malicious code detection, but these methods also rely on specialized knowledge and custom rules.

Recently, with the rise of deep learning techniques, learning-based malicious code detection methods have become increasingly popular. Various deep learning and machine learning methods have been employed for malicious code detection. Among them, Yadav et al. \citep{DBLP:journals/compsec/YadavMRVP22} converts bytecode into images and uses convolutional neural networks (CNN) for classification, achieving high detection accuracy. They also applies a pseudo-label stacked autoencoder for semi-supervised learning, improving the model's generalization capability. At the same time, many studies focus on extracting different types of features (such as API calls, permissions, system calls, opcode, etc.) for malicious code detection, thus expanding the application range of learning-based methods. Muhammad et al. \citep{2024AMDDLmodel} extracts opcode frequencies, API calls, and permissions, transforms them into two-dimensional images, and employs convolutional neural networks (CNN) for malicious code detection. Wu et al. \citep{DBLP:journals/iet-ifs/WuSWZS23} utilizes text mining methods to extract key features from application code, followed by the generation of call graphs and the integration of Bi-LSTM and GNN for analysis. Many other studies focus on enhancing the robustness of malicious code detection models against adversarial attacks. Yumlembam et al. \citep{DBLP:journals/iotj/YumlembamIJY23} combines variational graph autoencoders (VGAE) with generative adversarial networks (GAN) to strengthen the model's robustness when attackers are aware of the model's features. Amin et al. \citep{DBLP:journals/ett/AminSSAKA22} uses long short-term memory generative adversarial networks (LSTM-GAN) to process opcode sequence features, improving the model’s ability to detect adversarial attacks. The methods mentioned above are all based on learning from malicious code data and constructing models. Li et al. \citep{DBLP:conf/icse/Li00XLX024}, on the other hand, approaches the problem from the perspective of the attacker, successfully allowing malicious code variants to persist in machine learning-based, signature-based, and hybrid anti-malicious code software. Li et al. \citep{DBLP:conf/icse/Li00XLX024} utilizes the uncertainty estimation from the corrected model output to adjust the prediction results, thereby enhancing the accuracy of the DNN model. As for methods related to the recently popular large language models, we will introduce them in the next part.

\subsection{Malicious Code Detection within PyPI}

The widespread use of open-source software has made PyPI vulnerable to significant security challenges, with malicious code attacks leading to global disruptions and billions of dollars in losses. Okafor et al.  \citep{okafor2022sok} highlighted issues such as service interruptions and cybersecurity risks. Dambra et al. \citep{dambra2023decoding} noted that many studies have proposed machine learning models for malicious code detection, achieving near-perfect performance, but with differing methodologies and feature extraction techniques.

Static code analysis is a key technique for detecting malicious Python packages. Gobbi and Kinder \citep{gobbi2023using} proposed using the CodeQL framework for detecting malicious npm packages, successfully identifying 125 malicious packages with no false positives. Liu et al. \citep{liu2024eatvul} discussed challenges posed by machine learning-based malicious code detection that relies on binary files, presenting new opportunities for static analysis. Dambra et al. \citep{dambra2023decoding} collected the largest balanced dataset for malicious code detection, revealing that static features outperform dynamic ones, and that larger datasets and more samples per family improve accuracy. Arp et al. \citep{arp2022dos} emphasized the importance of dataset selection and evaluation in applying machine learning to security. Vu et al. \citep{DBLP:conf/icse/VuNM23} conduct study that reveal repository administrators require extremely low false positive rates (below 0.1\%). They found that a socio-technical malicious code detection system has emerged, where external security researchers scan for malicious code, filter the results, and report to administrators. MPHunter \citep{DBLP:conf/kbse/LiangLWLW23} and Ea4mp \citep{DBLP:conf/kbse/0001GCB0024} introduced language model into malicious code detection.

Malicious code detection within Python packages faces challenges such as anti-detection techniques like obfuscation, limited review capabilities of maintainers, and high false positive rates. Future research should focus on hybrid detection methods combining static analysis, machine learning, and behavior analysis, alongside advancing techniques to counteract code obfuscation and exploring the use of large language models in detection.

\subsection{LLM for Malicious Code Detection}
With the rapid development of Large Language Models (LLMs), their application in the field of malicious code detection, particularly in code analysis, behavior analysis, data augmentation, malicious code generation, and integrated frameworks, has garnered widespread attention.

LLMs leverage their powerful code comprehension capabilities to analyze source code or decompiled code to identify potential malicious behaviors. Research by Fang et al. \citep{fang2024llm} indicates that while LLMs perform well in certain code analysis tasks, their performance significantly declines when dealing with complex or obfuscated code. The Maltracker method proposed by Yu et al. \citep{yu2024maltracker}, which combines LLMs with traditional static analysis techniques, significantly improves the accuracy of malicious code detection. In contrast to static code analysis, behavior-based methods focus on the dynamic behavior of malicious code during execution. Akinsowon and Jiang \citep{akinsowon2024leveraging} demonstrate that LLMs can significantly improve detection rates by analyzing the dynamic behavioral characteristics of malicious code, especially for those malicious code samples that are difficult to detect through static analysis.

The generative capabilities of LLMs provide solutions to the issue of insufficient malicious code samples. Yu et al. \citep{yu2024maltracker} extend the diversity and representativeness of datasets by translating malicious functions into JavaScript, thus enhancing the accuracy of malicious code detection. The powerful generative ability of LLMs can both enhance security defenses and potentially be misused to generate more sophisticated malicious software. Khan et al. \citep{khan2024exploring} propose a risk mitigation framework aimed at preventing malicious code through the use of LLMs.

Given the complexity of evolving malicious code threats, single technologies are often inadequate. The integrated framework proposed by Khan et al. \citep{khan2024exploring} combines the advantages of LLMs with traditional security techniques, designing a multi-layered, multi-module detection system that effectively responds to constantly changing malicious code threats.

\subsection{Distinctions from Code Embedding and Dynamic Analysis Methods}

Beyond metadata- and rule-based detectors, recent work has explored pre-trained code models and dynamic traffic analysis for software security. However, these approaches differ from our setting in both granularity and analysis paradigm.

Several recent methods build security detectors on top of CodeBERT. For example, ugSliceVul~\citep{zou2025code} and VulCoBERT~\citep{Xia2024VulCoBERT} utilize pre-trained models for function-level vulnerability detection, and typically operate on short C/C++ snippets. Likewise, the Multiclass Software Defect Prediction (MSDP) model by Hussain et al.~\citep{hussain2025leveraging} and VulD-CodeBERT by Xiong and Dong~\citep{xiong2024vuld} demonstrate that fine-tuned CodeBERT can effectively classify diverse defects and vulnerabilities at the snippet or file level. In contrast, malicious Python packages in our setting often rely on project-wide logic spanning multiple modules and installation hooks (e.g., \texttt{setup.py}). The standard context window of these models (e.g., 512 tokens for CodeBERT~\citep{feng2020codebert}) forces heavy truncation when applied to entire packages with thousands of tokens, making it difficult to capture long-range, cross-file interactions. Although some studies apply CodeBERT to Python source snippets \citep{Zhao2024PythonCodeBERT}, to the best of our knowledge they do not release reproducible, end-to-end systems for package-level scanning. For this reason, we adopt MPHunter and Ea4mp as our primary baselines, as they are specifically tailored to package-level Python/PyPI malware.

Orthogonal to static code analysis, MalPaCA~\citep{nadeem2019malpaca} focuses on clustering malware based on network traffic traces captured during sandboxed execution. This represents a dynamic analysis paradigm that requires executing each sample and collecting PCAP logs. Our GMLLM framework, by contrast, targets \emph{static} supply-chain scanning: it operates directly on Python source code and dependency/call graphs, aiming to detect threats before installation or execution. Using MalPaCA as a baseline in our setting would require dynamically executing thousands of packages to generate traffic traces, which is impractical at PyPI scale and conceptually distinct from our static-analysis objective.

\section{Implementation Details}
\subsection{Graph Construction Details}
\label{apx:GCD}
We begin the graph construction by extracting the dependencies within the code. We traverse all `.py` files in the project directory and parse the source code of each file into an AST object, constructing the set $\mathcal{V}^{\text{code}}$ based on the nodes in the AST. The $\mathcal{V}^{\text{code}}$ includes class and function nodes within the AST, and each Python file itself is treated as a "module" type node. The source code corresponding to each node will be stored as a node attribute.

The edge set $\mathcal{E}^{\text{code}}$ is divided into dependency edges and call relationship edges. The dependency edges include definition edges, inheritance edges, and decorator edges, which are obtained through AST parsing. These edges are derived as follows:

1. \textbf{Definition Edge}: When a definition is encountered within a scope, a definition edge is created from the current scope's node to the newly defined class or function node.

2. \textbf{Inheritance Edge}: When processing class definition nodes, we check the base classes and create an inheritance edge from the current class node to each of its base class nodes.

3. \textbf{Decorator Edge}: We iterate over the decorator list for class and function definitions. By parsing the decorator expressions, it creates a decorator edge from the function node representing the decorator to the node representing the decorated class or function.

Next, we outline the call relationship edges. We traverse the AST again to identify and record these relationships. We specifically target the AST nodes representing function calls to identify function invocations. To accurately resolve the identity of the called function, the analyzer combines various pieces of information, including the import mapping, aliases, and the current scope context. Once the function is identified, an edge is created from the calling node to the called function node. These call relationships are recorded as different types of edges based on the context, including:

1. \textbf{Function-level Call}: If the invocation occurs within another function or method, a Function-level Call edge is recorded.

2. \textbf{Module-level Call}: If the invocation occurs at the module level (i.e., outside any function or class), a Module-level Call edge is recorded.

3. \textbf{Hook}: We specifically handles calls to the `setup()` function or `setuptools.setup()` function. We check the `cmdclass` parameter, and if a custom class overrides the installation command (`install`), a hook edge is created from the `setup()` node to the `run()` method of the custom class, capturing this potentially high-risk installation-time code execution behavior.

\subsection{Rule Generation Details}
\label{apx:rgd}

As for the construction of the each rule $s$ within $\mathcal{S}$, we used Python's lambda expressions and employed a functional programming approach to dynamically match different code behaviors. Each rule determines whether a behavior matches a malicious characteristic based on function names, method calls, or module names. The following presents two examples:

\begin{tcolorbox}[
    colback=codebg,     
    colframe=black,     
    boxrule=0.5pt,      
    arc=4pt,            
    title=Examples of Sensitive Behavior Rules, 
    fonttitle=\bfseries,
    fontupper=\ttfamily
]
\textcolor[RGB]{0,128,0}{"network"}: \textcolor[RGB]{0,0,255}{lambda} n: n.startswith((\textcolor[RGB]{0,128,0}{"socket."}, \textcolor[RGB]{0,128,0}{"requests."}, \textcolor[RGB]{0,128,0}{"urllib."})), 

\textcolor[RGB]{0,128,0}{"phishing"}: \textcolor[RGB]{0,0,255}{lambda} n: n in (\textcolor[RGB]{0,128,0}{"requests.post"}, \textcolor[RGB]{0,128,0}{"HTTPConnection"})
\end{tcolorbox}

The 'network' rule checks if network-related libraries (such as socket, requests, urllib) are being invoked, while the 'phishing' rule focuses on HTTP request methods potentially related to phishing attacks. The rules are defined as conditional functions through lambda expressions, returning True when specific conditions are met, indicating that the behavior matches the malicious characteristic. Compared to other approaches, we found that the automatic generation based on LLM (large language models) is more suitable for this simple and explicit rule construction pattern.

The prompt that is used to generate $\mathcal{S}^{\text{comm}}$ is as follows:

\begin{tcolorbox}[
    colback=red!10,            
    colframe=DarkRed,      
    boxrule=0.5pt,             
    arc=4pt,                   
    title=Prompt $\mathcal{S}^{\text{comm}}$, 
    fonttitle=\bfseries
]
\{ ``role": ``system", 

``content": ``````You are an AI model designed to identify and analyze sensitive behaviors in Python programming code. Sensitive behaviors refer to actions or patterns in the code that could result in security vulnerabilities, data privacy violations, or unintended exposure of sensitive information. These behaviors include but are not limited to insecure network operations, improper handling of user inputs, data encryption weaknesses, and any practices that may inadvertently expose sensitive data or increase the risk of attacks.

\quad Please provide a list of potential sensitive behaviors based upon common Python coding and analysis knowledge. For each sensitive behavior, provide a brief explanation of why it is considered sensitive and how it could lead to vulnerabilities or privacy violations. Additionally, provide a Python function (using lambda expressions and a functional programming approach) to detect and match the corresponding behavior. The function should analyze elements like function names, method calls, variable names, or imported modules.

\quad The rule for each sensitive behavior should determine whether a particular piece of code exhibits a malicious or risky characteristic based on its structure, such as the use of certain function names, method calls, or external libraries. Here are some examples:

\quad Example 1: Detecting Network Operations:

\quad ``network": lambda n: n.startswith(``socket.", ``requests.", ``urllib."),

\quad ......

\quad Each rule should be a lambda expression that matches specific patterns, function calls, or method names relevant to the corresponding sensitive behavior.""" \}

\end{tcolorbox}

The prompt that is used to generate $\mathcal{S}^{\text{data}}$ is as follows:

\begin{tcolorbox}[
    colback=red!10,            
    colframe=DarkRed,      
    boxrule=0.5pt,             
    arc=4pt,                   
    title=Prompt $\mathcal{S}^{\text{data}}$, 
    fonttitle=\bfseries
]
\{ ``role": ``system", 

``content": ``````You are an AI model designed to identify and analyze sensitive behaviors in Python programming code. Sensitive behaviors refer to actions or patterns in the code that could result in security vulnerabilities, data privacy violations, or unintended exposure of sensitive information. These behaviors include but are not limited to insecure network operations, improper handling of user inputs, data encryption weaknesses, and any practices that may inadvertently expose sensitive data or increase the risk of attacks.

\quad You are provided with a Python code snippet that may contain sensitive behaviors or security vulnerabilities. Please analyze the code and summarize, and provide a list of potential sensitive behaviors detected in the provided Python code. For each sensitive behavior, provide a brief explanation of why it is considered sensitive and how it could lead to vulnerabilities or privacy violations. Additionally, provide a Python function (using lambda expressions and a functional programming approach) to detect and match the corresponding behavior. The function should analyze elements like function names, method calls, variable names, or imported modules.

\quad The rule for each sensitive behavior should determine whether a particular piece of code exhibits a malicious or risky characteristic based on its structure, such as the use of certain function names, method calls, or external libraries. Here are some examples:

\quad Example 1: Detecting Network Operations:

\quad ``network": lambda n: n.startswith(``socket.", ``requests.", ``urllib."),

\quad ......

\quad Each rule should be a lambda expression that matches specific patterns, function calls, or method names relevant to the corresponding sensitive behavior.""" 
\}

            \{``role": ``user", ``content": \textbf{\textcolor{blue}{$<X_{i}>$}}\}

\end{tcolorbox}

\subsection{Prompt $\rho^{\text{ana}}$}
\label{apx:ana}
The detailed content concerning $\rho^{\text{ana}}$ is:
\begin{tcolorbox}[
    colback=red!10,            
    colframe=DarkRed,      
    boxrule=0.5pt,             
    arc=4pt,                   
    title=Prompt $\rho^{\text{ana}}$, 
    fonttitle=\bfseries
]
            \{ ``role": ``system", 
            
            ``content": 
            ``````You are a PyPI package security auditor. 
            You have been provided with the 'high-attention subgraph' structure of a PyPI package script, 
            which only includes node names and call relationships. Based solely on this subgraph structure, please answer the following:
            Is this structure indicative of potential malicious activity? (Respond only with 'Malicious' or 'Benign')
            
            Provide your reasoning. Be cautious not to label a package as malicious based on a single suspicious behavior without considering the broader context of the entire report. If mitigation is needed, identify the highest priority nodes or calls for input validation or permission checks. Response Format:

            \quad        Verdict:
                    
            \quad        Reasoning:
                    
            \quad        Mitigation:"""\},
                    
            \{``role": ``user", ``content": \textbf{\textcolor{blue}{$<\text{Att}(G^{\text{code}}_{j})>$}}\}
\end{tcolorbox}

\subsection{Details Concerning Evaluation Upon Malicious Behavior Description}
\label{apx:judge}
The evaluation was conducted exclusively on samples drawn from the malicious subset of the MalCP dataset, each of which includes a verified ground-truth behavior summary from Synk~\citep{snyk_vuln_db}.

We selected GPT-4o as the evaluator due to its superior reasoning and instruction-following capabilities among all models tested in our experiments. Each explanation-ground truth pair was fed into the model using the same prompt template, and results were collected via automated scripts.

For models that misclassified these malicious samples as benign, a zero score was assigned across all evaluation dimensions as a penalty.
This penalty mechanism is justified as models that fail to recognize malicious intent inherently cannot provide meaningful behavioral explanations.  The full prompt template is shown below.
\begin{tcolorbox}[
    colback=red!10,            
    colframe=DarkRed,      
    boxrule=0.5pt,             
    arc=4pt,                   
    title=Prompt $\rho^{\text{judge}}$, 
    fonttitle=\bfseries
]
            \{ ``role": ``system", 
            
            ``content": 
            ``````You are a senior cybersecurity analyst with deep expertise in threat intelligence. Your role is to evaluate an AI-generated explanation of malicious behavior by its depth of insight, not just its factual accuracy.

Your assessment has two parts:

Part 1: Intrinsic Quality
Evaluate the explanation’s internal strength—without referencing the ground truth—using the following criteria:

1. Threat Tactic Generalization (1–5)

How well does it generalize from specific functions or code elements to a recognized cybersecurity tactic (e.g., `Reconnaissance', `Defense Evasion', `Exfiltration')?

→ 5: Clearly identifies and frames the analysis using a standard attack pattern (e.g., MITRE ATT\&CK).

→ 4: Identifies a general malicious purpose (e.g., data collection), but lacks formal tactical framing.

→ 3: Recognizes suspicious behavior but generalizes weakly or vaguely (e.g., ``could be malicious").

→ 2: Lists actions with minimal abstraction (e.g., ``gets username → collects info").

→ 1: Only describes concrete functions with no generalization.

2. Execution Path Traceability (1–5)

How clearly does it reconstruct a step-by-step execution flow, including function calls, data movement, and control logic?

→ 5: Presents a complete, logical, and verifiable sequence of actions.

→ 4: Shows a mostly coherent flow but omits minor steps or data dependencies.

→ 3: Outlines key stages (e.g., collect → encode → send), but with gaps or unclear transitions.

→ 2: Mentions multiple functions without clear order or causality.

→ 1: No discernible execution path.

3. Evidence Groundedness (1–5)

Are claims directly and tightly supported by specific code elements (e.g., function calls, strings, variables)?

→ 5: Every significant claim is explicitly tied to observable code.

→ 4: Most claims are evidence-backed; minor inferences are reasonable.

→ 3: Core behaviors are supported, but some conclusions extend beyond direct evidence.

→ 2: Sparse or generic references to code; many unsupported assertions.

→ 1: Claims are vague, speculative, or entirely disconnected from code.

Part 2: Factual Alignment

Compare the explanation to the Ground Truth Summary:

4. Factual Alignment (1-5)

Does it correctly identify and accurately describe the primary malicious behavior?

→ 5: Accurate and complete—captures the core behavior and key details.

→ 3: Partially correct—identifies the general type of malicious activity but misses or misrepresents critical elements.

→ 1: Fundamentally wrong or entirely misses the core behavior.

 OUTPUT (JSON):

\{

              \quad        ``quality\textunderscore scores": \{
  
              \quad        ``threat\textunderscore tactic\textunderscore generalization": int,
    
              \quad        ``execution\textunderscore path\textunderscore traceability": int,
    
              \quad        ``evidence\textunderscore groundedness": int
    
  \},
  
  ``alignment\textunderscore score": \{
  
              \quad        ``factual\textunderscore alignment": int
    
  \}
\}
"""\},
                    
            \{``role": ``user", ``content": Explanation\textunderscore To\textunderscore Evaluate:
              \quad       ``\{explanation\}"

Ground\textunderscore Truth\textunderscore Summary:  
              \quad       ``\{ground\textunderscore truth\}"\}
\end{tcolorbox}

The model returns structured JSON output, including both quantitative scores and a textual justification. To ensure response validity, all outputs were automatically checked for schema compliance (valid JSON).  If an output was malformed, we re-issued the query up to 3 times.  Only consistently valid outputs were included in final scoring;  invalid ones were excluded from aggregate metrics. A concrete example is given below:

\begin{tcolorbox}[
    colback=codebg,     
    colframe=black,     
    boxrule=0.5pt,      
    arc=4pt,            
    title=LLM-as-a-Judge Evaluation Examples, 
    fonttitle=\bfseries,
    fontupper=\ttfamily
]
\textcolor[RGB]{0,128,0}{"quality\textunderscore scores"}:

\{\textcolor[RGB]{0,128,0}{"threat\textunderscore tactic\textunderscore generalization"}: 4, \textcolor[RGB]{0,128,0}{"execution\textunderscore path\textunderscore traceability"}: 3, \textcolor[RGB]{0,128,0}{"evidence\textunderscore groundedness"}: 3\},

\textcolor[RGB]{0,128,0}{"alignment\textunderscore score"}: \{\textcolor[RGB]{0,128,0}{"factual\textunderscore alignment"}: 3\}, 

\textcolor[RGB]{0,128,0}{"Reasoning"}: "The provided Python code is a sophisticated malware that performs a variety of malicious activities. It is designed to steal sensitive information such as passwords, cookies, and tokens from various applications and browsers. The code includes functionality to decrypt encrypted data, exfiltrate data to a remote server, and manipulate clipboard data to replace cryptocurrency wallet addresses. It also attempts to evade detection by checking for virtual machine environments and debugging tools. Additionally, the code includes a mechanism to persistently install itself on the victim's system by adding itself to the startup programs. The use of obfuscation and deceptive descriptions further indicates malicious intent. The code interacts with external servers using hardcoded URLs, which are used to send stolen data and receive commands. This behavior is typical of malware designed for data theft and system compromise."

\end{tcolorbox}

\subsection{Human Validation of Explanation-Quality Metrics}
\label{apx:humanjudge}
To validate the reliability of using GPT-4o as an automatic evaluator for the explanation-quality metrics in Table~\ref{tab:description}, we conducted a human study with two security practitioners.

We randomly sampled 200 model-generated explanations from MalCP. For each explanation, the two experts independently scored four dimensions using the same discrete scale as in the main paper: \emph{Threat Tactic Generalization}, \emph{Execution Path Traceability}, \emph{Evidence Groundedness}, and \emph{Factual Alignment}. We then computed (i) inter-annotator agreement between the two experts and (ii) the correlation between GPT-4o’s scores and the mean of the two human scores.

Table~\ref{tab:human_validation_results} reports Spearman’s rank correlation coefficients ($\rho$) for both Human–Human and GPT-4o–Human comparisons. Human–Human agreement is consistently moderate to strong ($\rho \in [0.63, 0.81]$, $p < 10^{-7}$), which is in line with the natural variability expected for this kind of subjective, multi-dimensional task. GPT-4o’s scores exhibit similarly strong correlations with the human averages ($\rho \in [0.72, 0.88]$, $p < 10^{-11}$) across all four dimensions, in some cases slightly higher than Human–Human.

\begin{table}[htbp]
    \centering
    \caption{Spearman’s $\rho$ (and $p$-values) for Human--Human agreement and GPT-4o--Human correlation on the four explanation-quality dimensions.}
    \label{tab:human_validation_results}
    \resizebox{\textwidth}{!}{
        \begin{tabular}{lcccc}
            \toprule
            \textbf{Evaluation Dimension} & \textbf{$\rho$ (Human--Human)} & \textbf{p-value (Human--Human)} & \textbf{$\rho$ (GPT--Human)} & \textbf{p-value (GPT--Human)} \\
            \midrule
            \texttt{Threat Generalization} & 0.630 & $3.68 \times 10^{-8}$ & 0.815 & $5.57 \times 10^{-16}$ \\
            \texttt{Execution Path Traceability}  & 0.802 & $1.69 \times 10^{-8}$ & 0.879 & $1.49 \times 10^{-16}$ \\
            \texttt{Evidence Groundedness}         & 0.749 & $4.61 \times 10^{-9}$ & 0.755 & $3.02 \times 10^{-14}$ \\
            \texttt{Factual Alignment}             & 0.702 & $6.60 \times 10^{-12}$ & 0.729 & $2.69 \times 10^{-12}$ \\
            \bottomrule
        \end{tabular}
    }
\end{table}

Overall, these results support the use of GPT-4o as a reasonable proxy for expert judgment when evaluating explanation quality, while the core \emph{Factual Alignment} metric in the main paper already relies directly on expert-curated malware analysis reports as ground truth.

\section{Experiment Details}

\subsection{Datasets}
\label{apx:dataset}
To evaluate our method under real-world conditions, we constructed a benchmark dataset named \textbf{MalCP (Malicious Codes from PyPI)}. This dataset integrates malicious packages from three public sources, aligns their behavioral labels using the Snyk advisory system~\citep{snyk_vuln_db}, and filters them using size-based heuristics to ensure reproducibility and label consistency.

\subsubsection{Data Aggregation from Public Sources}

To construct the MalCP dataset, we compiled malicious Python packages from the PyPI ecosystem by integrating multiple open-source intelligence sources, including Mal\_OSS~\citep{DBLP:conf/kbse/GuoXLHFL23}, Backstabbers Knife Collection~\citep{ohm2020backstabber}, and Datadog's open malware dataset~\citep{guarddog2023malicious}. This resulted in a large and diverse collection of real-world malicious packages.

While the source datasets provide access to source code and limited metadata (such as upload timestamps or package descriptions), the vast majority do not contain structured or human-readable descriptions of malicious behaviors. To enable behavior-aware detection and analysis, we retrieved and matched advisory content from the \textbf{Snyk Vulnerability Database}.

Specifically, for each package in our initial collection, we attempted to crawl the \texttt{overview} section from the Snyk Vulnerability Database, which typically contains a concise, high-level description of the malicious activity. Successful matches resulted in 7,377 annotated packages. These overviews were stored in a structured \texttt{vuln\textunderscore report.json} file for each sample. Examples include:

\begin{quote}
``10Cent11 is a malicious package. It creates a reverse shell to a hard-coded IP address, giving the attacker full control over an infected system.''
\end{quote}

This standardization ensures that all labels are derived from a consistent and widely trusted security knowledge base.

\subsubsection{Size-Based Filtering Criteria}

To ensure the analytical value and structural completeness of the dataset, we applied a \textbf{size-based filtering policy} that eliminated trivial or non-informative samples while preserving diversity across package sizes. The thresholds of \textbf{5KB} and \textbf{10KB} were empirically determined by inspecting the distribution of raw package sizes:

\begin{itemize}
  \item \textbf{Large} Packages larger than 10KB: Fully retained.
  \item \textbf{Medium} Packages between 5KB and 10KB: Partially sampled.  We randomly sampled a representative subset to maintain approximate proportionality with the large-size group. Sampling was performed using a fixed random seed for reproducibility.
  \item \textbf{Small} Packages smaller than 5KB: A small, balanced subset. We retained a small, behaviorally diverse subset. At least one sample was selected per behavior category (e.g., reverse shell, credential theft), ensuring semantic diversity despite the size limitation.
\end{itemize}

The distribution of the final 1,659 packages by size category is shown in Table~\ref{tab:size-distribution}.

\begin{table}[h]
\centering
\caption{Final Samples by Package Size}
\label{tab:size-distribution}
\begin{small}
\begin{tabular}{lcc}
\toprule
\textbf{Package Size Range} & \textbf{Samples Retained} & \textbf{Selection Criteria} \\
\midrule
Large(\textgreater{} 10 KB) & 461 & All retained \\
Medium(5--10 KB)             & 588   & Partial sampling \\
Small(\textless{} 5 KB)     & 610   & Balanced selection \\
\midrule
\textbf{Total}       & 1,659 & -- \\
\bottomrule
\end{tabular}
\end{small}
\end{table}

\subsubsection{Attack-Type Distribution in MalCP}
\label{apx:}

For each malicious package in MalCP, we annotate its primary attack type by combining
(1) the human-written analysis reports from Snyk (used as ground-truth labels) and 
(2) the high-level tactics in the MITRE ATT\&CK framework \cite{strom2018mitre}. 
We collapse the original fine-grained descriptions into eight coarse-grained categories:
Execution, Impact, Credential Access / Collection, Defense Evasion, Exfiltration, 
Command and Control, Persistence, and Other (for rare or mixed behaviors).

Table~\ref{tab:malcp_attack_types} summarizes the resulting distribution. 
Execution covers cases where the package directly launches or injects a payload 
(e.g., malicious installation hooks); 
Impact captures destructive or resource-abuse behaviors (e.g., wiping files, crypto-mining);
Credential Access / Collection includes stealing API keys, SSH keys or browser credentials;
the remaining categories correspond to evasion, data exfiltration, command-and-control 
channels, and long-lived persistence mechanisms.

As the table shows, the three most common tactics---Execution (33.8\%), Impact (23.1\%), 
and Credential Access / Collection (23.0\%)---each account for roughly one third of 
malicious samples, while the remaining tactics form a non-trivial tail 
(Defense Evasion 8.0\%, Exfiltration 6.4\%, Command and Control 2.5\%, Persistence 0.2\%, 
Other 3.0\%). 
This reflects realistic attack behavior observed in the wild rather than an artificially 
balanced benchmark, and ensures that MalCP exercises detectors across a broad range of 
malicious intents.

\begin{table}[htbp]
\centering
\caption{Distribution of Malicious Samples across MITRE ATT\&CK Tactics}
\label{tab:malcp_attack_types}
\begin{small}
\begin{tabular}{lrr}
\toprule
Attack Type & Count & Percentage \\
\midrule
Execution                         & 561 & 33.82 \\
Impact                            & 383 & 23.09 \\
Credential Access / Collection    & 382 & 23.03 \\
Defense Evasion                   & 132 &  7.96 \\
Exfiltration                      & 106 &  6.39 \\
Other                             &  49 &  2.95 \\
Command and Control               &  42 &  2.53 \\
Persistence                       &   4 &  0.24 \\
\midrule
Total                             & 1659 & 100 \\
\bottomrule
\end{tabular}
\end{small}
\end{table}

\subsubsection{Dataset Filtering Pipeline}

The construction of the MalCP dataset followed a three-stage process:

\begin{enumerate}
  \item \textbf{Aggregation:} Malicious packages were consolidated from the PyPI ecosystem using multiple open-source intelligence sources.
  \item \textbf{Behavioral Annotation:} Each package was enriched with a structured behavior summary derived from the Snyk advisory system.
  \item \textbf{Filtering:} Samples were filtered based on archive structure and size constraints to ensure semantic richness and practical utility.
\end{enumerate}

\subsection{Settings}
\label{apx:setting}
This section outlines the end-to-end pipeline used in our experiments, including graph preparation, model training, explanation generation, and runtime environment setup.

\subsubsection{Data Splitting and Leakage Prevention}
\label{app:data-split}

For all benchmark datasets, we use package-level train/test splits with an 80/20 ratio. The package, rather than an individual file, function, or graph node, is treated as the atomic unit of splitting. This prevents code fragments from the same package from appearing in both the training and test sets.

The split is performed independently for each benchmark dataset. For datasets containing both malicious and benign packages, we preserve the class distribution across the train and test splits whenever possible. For MalCP, the size categories (Large, Medium, and Small) are used for reporting and analysis, and the package-level split is applied before computing the corresponding size-bucket results.

Before splitting, we remove duplicated and near-duplicated packages to reduce leakage across data sources. In particular, when constructing MalCP from multiple sources, we deduplicate packages using package metadata and normalized source-code fingerprints. Packages with identical names and versions, identical archive hashes, or highly similar normalized code contents are merged or removed before the train/test split.

For GMLLM, all learned or data-derived components are constructed exclusively from the training split. This includes the GNN parameters and the data-derived sensitive behavior rules $S^{data}$. The held-out test split is used only for evaluation. Baselines that do not require training, such as rule-based tools and direct LLM prompting, are evaluated on the same held-out test packages to ensure fair comparison.
We repeat the splitting and training process over five independent random seeds and report the average results in the main paper.

\subsubsection{Graph Preparation and Labeling}

Each package is parsed into a static code graph as described in \textbf{Appendix}~\ref{apx:GCD}. Nodes represent program elements (e.g., functions, classes, literals, or modules) and are enriched with metadata such as identifier name, AST type, and file-level context. Edges capture both internal control flow (e.g., call relationships) and external dependencies (e.g., inheritance or decoration).

Each node feature vector consists of:
\begin{itemize}
    \item a 64-dimensional embedding for the identifier name;
    \item a 16-dimensional embedding of the AST type;
    \item a one-hot vector representing malicious behavior labels.
\end{itemize}

Behavior labels are inferred from a synthesized rule set generated by prompting an LLM with structured behavior definitions. Each rule is a restricted lambda expression compiled via an AST whitelist. If rule synthesis fails, a fallback static rule set is used. Behavior features are stored as node attributes and aligned to a fixed \texttt{behavior2idx} vocabulary.

The graphs are then serialized into PyTorch Geometric-compatible tensors using frozen vocabularies (\texttt{name2idx}, \texttt{type2idx}, \texttt{edge\_type2idx}, \texttt{behavior2idx}) constructed from the training set and reused across all runs.

\subsubsection{Graph Classification (for Explanation Only)}

To extract semantically meaningful subgraphs for explanation, we train a lightweight graph neural classifier over the prepared graphs. The model is not used for standalone detection.

We adopt a two-layer GCN architecture, where each node's input is the concatenated feature vector described above. Each GCN layer is followed by a ReLU activation and dropout (rate = 0.6). A global mean pooling layer aggregates node embeddings to the graph level, which is then fed into a linear classifier with softmax output to predict package-level maliciousness.

The train/test split follows \textbf{Appendix}~\ref{app:data-split}. We use the Adam optimizer (learning rate $= 10^{-3}$, weight decay $= 10^{-3}$), a batch size of 128, and 100 training epochs.

\subsubsection{Explanation and LLM Review}

To support human-interpretable analysis, we apply a mask-based graph explainer to the trained classifier to obtain edge-level salience scores that indicate which code relations contribute most to the prediction. We then construct an attention subgraph by retaining the top-ranked edges (and their associated local context) under a fixed budget, resulting in a compact, high-signal representation of the original call graph.

These extracted subgraphs are submitted to a large language model for semantic judgment. Each LLM response returns a structured output containing a verdict, reasoning, and possible mitigation. This procedure allows us to incorporate symbolic reasoning and external knowledge into the evaluation of structural malicious patterns.

Our framework supports different LLM backends in practice, but we do not depend on any particular model configuration for core results.

\subsubsection{Runtime Environment and Reproducibility}

Our experiments are conducted on standard research hardware with NVIDIA GPUs ($\ge$16GB  VRAM). All code is implemented in Python 3.7+, using the following core libraries:
\begin{itemize}
  \item PyTorch 1.13.1 with CUDA 11.7
  \item PyTorch Geometric 2.3.1
  \item Scikit-learn 1.0.2
  \item OpenAI SDK $\ge$1.39.0
\end{itemize}
We fix random seeds across Python, NumPy, and PyTorch for reproducibility. Static vocabularies, synthesized rules, and all critical dependencies are released as part of our codebase. CUDA full determinism is enabled where applicable.
{%

\subsection{Token Calculation Details}
\label{apx:token_methodology}
To ensure reproducibility and accurate cost estimation, token counts were computed using model-specific pre-trained tokenizers, avoiding character-based heuristics. 
For OpenAI models (e.g., GPT-4o), we used \texttt{tiktoken} with the \texttt{o200k\_base} encoding; for open-source models (e.g., Llama-3), we employed the official Hugging Face tokenizers. 
This accounts for model-specific vocabulary as well as chat-template-mandated control tokens (e.g., \texttt{<|im\_start|>}).

The reported token counts reflect the full input context of a single-turn inference pass, including the system prompt, the user query (containing the extracted subgraph), and template-induced formatting tokens. 
No chunking or context accumulation was applied; each prediction is based on a single, holistic prompt to assess the model’s capacity to reason over the compressed graph representation.

\section{Extra Experimental Results}

\subsection{Runtime of the GNN-Explainer Step}
\label{apx:runtime}

We profile the cost of the per-package mask optimization used to extract attention subgraphs (Sec.~\ref{sec:method_explainer}). The driver script logs wall-clock time from launching \texttt{explainer\_main.py} to its termination. All experiments run on a single NVIDIA GPU (CUDA~11.7, cuDNN~8.5) with a fixed budget of 100 gradient steps per package. 

Table~\ref{tab:runtime_stats} summarizes the runtime distribution on MalCP (Small/Medium/Large buckets) and on the Datadog dataset. Across all subsets, the mean per-package runtime is around 10--11 seconds, the median is close to 9 seconds, and the 95th percentile ($P_{95}$) stays below roughly 20 seconds. The similar numbers across size buckets are largely due to a substantial constant overhead (Python startup, checkpoint loading, graph preparation), together with the fixed 100-step optimization budget; the part of the computation that scales with graph size is not dominant at this scale.

\begin{table}[htbp]
    \centering
    \caption{Runtime statistics of the GNN-Explainer. 
    ``Small/Medium/Large'' are the MalCP size buckets; Datadog is one of the public datasets used in the main experiments.}
    \label{tab:runtime_stats}
    \begin{small}
    \begin{tabular}{lccccc}
        \toprule
        \textbf{Subset / Dataset} & \textbf{Mean $\pm$ Std (s)} & \textbf{Median (s)} & \textbf{Min / Max (s)} & \textbf{$P_{95}$ (s)} & \textbf{Exceeded $P_{95}$ (\%)} \\
        \midrule
        Small   & 10.18 $\pm$ 4.10 & 8.95 & 6.96 / 25.31 & 18.32 & 5.09 \\
        Medium  & 10.59 $\pm$ 3.92 & 8.96 & 7.08 / 25.47 & 18.57 & 5.05 \\
        Large   & 10.73 $\pm$ 4.70 & 8.86 & 6.83 / 55.28 & 19.16 & 5.04 \\
        Datadog & 10.76 $\pm$ 4.21 & 8.68 & 7.15 / 22.86 & 20.45 & 5.04 \\
        \bottomrule
    \end{tabular}
    \end{small}
\end{table}

Overall, the GNN-Explainer step adds roughly 10 seconds of offline processing per package under our fixed 100-step budget, with about 95\% of packages finishing within 18--20 seconds across different datasets. Since GMLLM is intended for \emph{offline batch scanning} of new releases, rather than interactive request-time serving, this level of overhead is acceptable, especially when combined with the
substantial reduction in LLM token usage reported in Sec.~\ref{sec:token}. In practice, the explainer step budget (currently 100) can be reduced (e.g., to 30--50 steps) or explanations can be cached across scans to further reduce latency.

\subsection{Evaluation on Additional LLM Backbones}
\label{apx:recent-llms}

We further evaluate GMLLM with two additional LLM backbones, Qwen3.5-Flash and DeepSeek-V3.2, on a stratified subset of 600 MalCP packages covering different labels and package sizes. Each backbone is compared under direct prompting and the GMLLM-enhanced setting. Table~\ref{tab:recent_llm_perf} reports the detection performance. GMLLM improves both large-package and overall performance for both additional backbones.

Table~\ref{tab:recent_llm_tokens} reports the corresponding average input token counts across package sizes. GMLLM substantially reduces token usage, especially on large packages. These results indicate that the benefit of GMLLM is not tied to the specific LLM backbones used in the main experiments.

\begin{table}[t]
\centering
\caption{Performance with additional LLM backbones.}
\label{tab:recent_llm_perf}
\begin{small}
\begin{tabular}{lcccc}
\toprule
Model & Large Recall & Large F1 & Overall Recall & Overall F1 \\
\midrule
Qwen3.5-Flash & 80.00 & 88.64 & 92.50 & 96.02 \\
GMLLM(Qwen3.5-Flash based)& \textbf{93.00} & \textbf{95.14} & \textbf{97.17} & \textbf{97.82} \\
DeepSeek-V3.2 & 75.50 & 82.29 & 90.00 & 93.43 \\
GMLLM(DeepSeek-V3.2 based) & \textbf{91.00} & \textbf{94.30} & \textbf{96.17} & \textbf{97.22} \\
\bottomrule
\end{tabular}
\end{small}
\end{table}

\begin{table}[t]
\centering
\caption{Average token count with additional LLM backbones across different package sizes.}
\label{tab:recent_llm_tokens}
\begin{small}
\begin{tabular}{lccc}
\toprule
Model & Large Tokens & Medium Tokens & Small Tokens \\
\midrule
Qwen3.5-Flash & 319,772 & 29,874 & 758 \\
GMLLM(Qwen3.5-Flash based) & \textbf{639} & \textbf{429} & \textbf{260} \\
DeepSeek-V3.2 & 324,202 & 29,352 & 714 \\
GMLLM(DeepSeek-V3.2 based) & \textbf{661} & \textbf{444} & \textbf{251} \\
\bottomrule
\end{tabular}
\end{small}
\end{table}
}

\subsection{Extra Visualization Results}

\begin{figure}
  \centering
  \begin{minipage}{0.4\textwidth}
    \includegraphics[width=\linewidth]{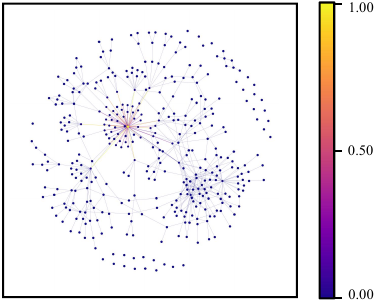}
    \subcaption{}
  \end{minipage}
  \begin{minipage}{0.4\textwidth}
    \includegraphics[width=\linewidth]{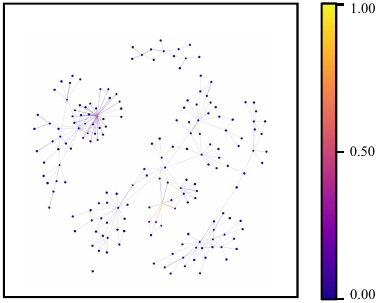}
    \subcaption{} 
  \end{minipage}
  \begin{minipage}{0.4\textwidth}
    \includegraphics[width=\linewidth]{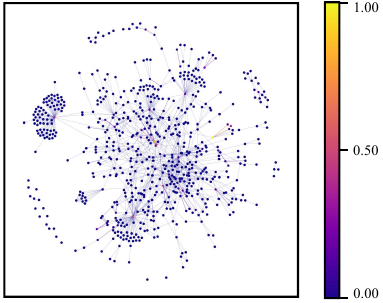}
    \subcaption{} 
  \end{minipage}
    \begin{minipage}{0.4\textwidth}
    \includegraphics[width=\linewidth]{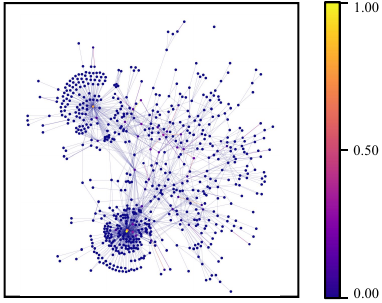}
    \subcaption{} 
  \end{minipage}
  \caption{Visualizations of the code graph attention outputted with GMLLM when processing programs of large scales.}
  \label{fig:el} 
\end{figure}
\begin{figure}
  \centering
  \begin{minipage}{0.4\textwidth}
    \includegraphics[width=\linewidth]{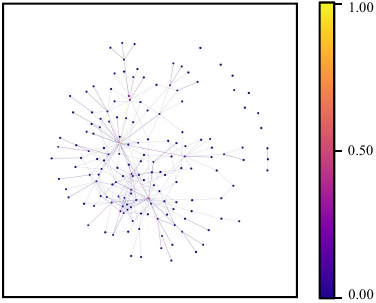}
    \subcaption{}
  \end{minipage}
  \begin{minipage}{0.4\textwidth}
    \includegraphics[width=\linewidth]{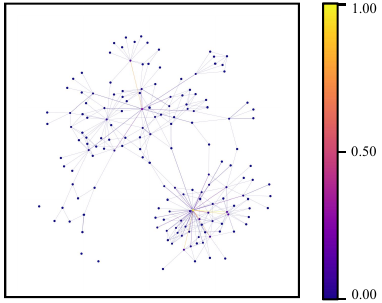}
    \subcaption{} 
  \end{minipage}
  \begin{minipage}{0.4\textwidth}
    \includegraphics[width=\linewidth]{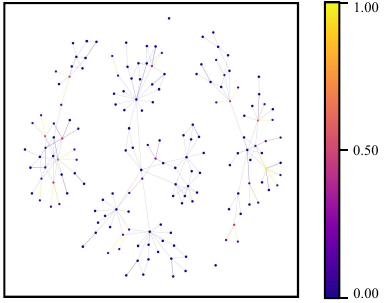}
    \subcaption{} 
  \end{minipage}
    \begin{minipage}{0.4\textwidth}
    \includegraphics[width=\linewidth]{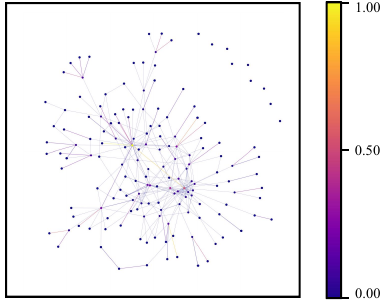}
    \subcaption{} 
  \end{minipage}
  \caption{Visualizations of the code graph attention outputted with GMLLM when processing programs of medium scales.}
  \label{fig:em} 
\end{figure}
\begin{figure}
  \centering
  \begin{minipage}{0.4\textwidth}
    \includegraphics[width=\linewidth]{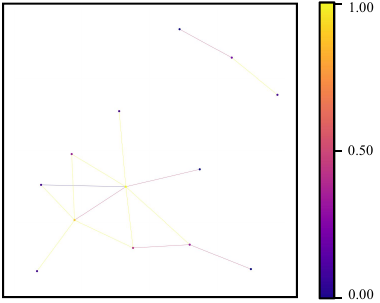}
    \subcaption{}
  \end{minipage}
  \begin{minipage}{0.4\textwidth}
    \includegraphics[width=\linewidth]{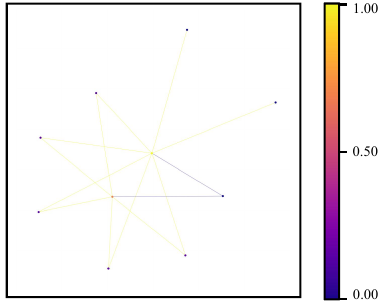}
    \subcaption{} 
  \end{minipage}
  \begin{minipage}{0.4\textwidth}
    \includegraphics[width=\linewidth]{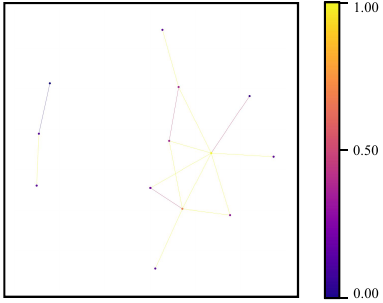}
    \subcaption{} 
  \end{minipage}
    \begin{minipage}{0.4\textwidth}
    \includegraphics[width=\linewidth]{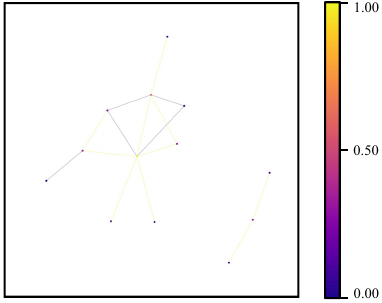}
    \subcaption{} 
  \end{minipage}
  \caption{Visualizations of the code graph attention outputted with GMLLM when processing programs of small scales.}
  \label{fig:es} 
\end{figure}

Figures \ref{fig:el}, \ref{fig:em}, and \ref{fig:es} provide additional visualizations of the code graph attention. From the figures, it can be observed that malicious code nodes or relationships, i.e., those nodes and edges with colors closer to yellow, often exhibit positions with a higher number of neighboring nodes or more complex associations. This topological property is quite evident and also suggests that graph representation learning is effective for the preliminary identification of these nodes.

\vfill\pagebreak
\subsection{Sample Model Judgments from High-Attention Subgraphs}
To further illustrate the security reasoning capability of our approach, we present illustrations of outputs of GMLLM. 

The following examples were generated by GMLLM based on ChatGPT 4o. For each case, the model produced a binary verdict (`malicious' or `benign'), a supporting explanation (`reason'), and a recommended mitigation (`mitigation').

These results demonstrate that our method enables the model to make accurate security judgments and articulate meaningful defenses—highlighting its potential for fine-grained, explainable threat analysis.

\vfill\pagebreak
\begin{tcolorbox}[
    colback=codebg,     
    colframe=black,     
    boxrule=0.5pt,      
    arc=4pt,            
    title= Output Example 1 
]
\{\textcolor[RGB]{0,0,0}{``name"}: ``hypedrop", \textcolor[RGB]{0,0,0}{``verdict"}: ``Malicious",
        
        \textcolor[RGB]{0,0,0}{``reasoning"}: ``The subgraph for this package contains numerous activities and calls that raise significant red flags and suggest malicious intent:  
        
        1. Password Decryption: The use of 
        
        `ctypes.windll.crypt32.CryptUnprotectData' indicates an attempt to decrypt sensitive data, which is commonly used to scrape encrypted data like saved passwords.  
        
        2. Clipboard Hijacking: The use of `pyperclip.copy' in combination with `auto\textunderscore copy\textunderscore wallet.address\textunderscore swap' suggests the possibility of clipboard manipulation, potentially to replace wallet addresses for cryptocurrency theft.  
        
        3. Credential Harvesting: Calls like `os.getlogin', `winreg.OpenKey', and `winreg.QueryValueEx' indicate that the package may be gathering system and user information illegitimately.  
        
        4. Screen and Cookie Stealing: Nodes like `PIL.ImageGrab.grab' (screenshot capture) and `steal\textunderscore cookies' strongly imply the stealing of private user data.  
        
        5. Unauthorized Memory Operations: Calls to 
        
        `ctypes.cdll.msvcrt.memcpy' and `ctypes.windll.kernel32.LocalFree' raise concerns about memory manipulation, which is often associated with malicious payload deployment or data harvesting.  
        
        6. Network Activity: The usage of `urllib.request.Request' and `urllib.request.urlopen' under the function `LoadUrlib' may imply exfiltration of data or receiving commands from a remote server.  
        
        7. Potential Anti-Debugging: The function `NoDebugg.sp3cCheq' suggests the package may contain anti-debugging checks to evade detection by security tools or analysts.  
        
        8. Execution and Persistence: The presence of `subprocess.Popen', combined with suspicious function names like `ZipThings' and system-manipulative calls like `os.\textunderscore exit', hints at unauthorized script execution or persistence mechanisms. 
        
        These behaviors, when taken together, are highly indicative of malicious software designed for data theft, remote control, or system exploitation.",
        
        \textcolor[RGB]{0,0,0}{``mitigation"}: ``The following nodes/calls should be flagged as highest priority for input validation, permission checks, or removal:  
        
        1. `ctypes.windll.crypt32.CryptUnprotectData': Validate any use of sensitive data decryption and ensure it is necessary and authorized.  
        
        2. `pyperclip.copy': Restrict clipboard access to prevent modification of user data.  
        
        3. `PIL.ImageGrab.grab' and `steal\textunderscore cookies': Explicitly validate and limit access to screen capture and browser storage to prevent data theft.  
        
        4. `ctypes.cdll.msvcrt.memcpy' and `ctypes.windll.kernel32.LocalFree': Review all direct memory manipulation for potential abuse.  
        
        5. `urllib.request.Request' and `urllib.request.urlopen': Monitor outgoing network calls to prevent unauthorized data exfiltration or command-and-control communication.  
        
        6. `os.getlogin', `winreg.OpenKey', and `winreg.QueryValueEx': Validate and restrict access to user information and registry keys.  
        
        7. `NoDebugg' Functions: Investigate all anti-debugging mechanisms and ensure they do not attempt to evade legitimate security analyses.  
        
        8. `subprocess.Popen': Confirm that subprocess spawning is securely configured and does not execute unauthorized commands or scripts.  
        
        These mitigations are critical to address the highly suspicious nature of the subgraph behavior."
    \}            
\end{tcolorbox}

\begin{tcolorbox}[
    colback=codebg,     
    colframe=black,     
    boxrule=0.5pt,      
    arc=4pt,            
    title= Output Example 2
]
\{\textcolor[RGB]{0,0,0}{``name"}: ``pipcoloradds", \textcolor[RGB]{0,0,0}{``verdict"}: ``Malicious",
        
        \textcolor[RGB]{0,0,0}{``reasoning"}: ``The subgraph structure shows the use of multiple functions and operations that are indicative of malicious behavior. Specifically:
        
        -  Data Theft:  The presence of `PIL.ImageGrab.grab' and `pipcoloradds-1.0.0.pipcoloradds.\textunderscore \textunderscore init\textunderscore \textunderscore .bc\textunderscore initial\textunderscore func.steal\textunderscore screen' implies potential screen grabbing functionality.
        
        -  Token and Credential Stealing:  The use of 
        
        `ctypes.windll.crypt32.CryptUnprotectData' indicates an attempt to decrypt sensitive data such as stored browser or system credentials. Additionally, 
        
        `pipcoloradds-1.0.0.pipcoloradds.\textunderscore \textunderscore init\textunderscore \textunderscore .bc\textunderscore initial\textunderscore func.steal\textunderscore token' makes use of network calls (`requests.get') and file system traversal (`os.walk'), which may be aimed at stealing access tokens or other sensitive information.
        
        -  Clipboard Manipulation:  The usage of `pyperclip.copy' with `pipcoloradds-1.0.0.pipcoloradds.\textunderscore \textunderscore init\textunderscore \textunderscore .auto\textunderscore copy\textunderscore wallet.address\textunderscore swap' suggests the possibility of modifying clipboard data, a tactic typically used in cryptocurrency address hijacking attacks.
        
        -  Exfiltration of Data:  Calls to `httpx.post'
        
        (pipcoloradds-1.0.0.pipcoloradds.\textunderscore \textunderscore init\textunderscore \textunderscore .bc\textunderscore initial\textunderscore func.ping\textunderscore on\textunderscore running) may indicate exfiltration of stolen data to a remote server.
        
        -  Code Execution:  The presence of `exec' within `pipcoloradds-1.0.0.pipcoloradds.\textunderscore \textunderscore init\textunderscore \textunderscore ' raises serious concerns as it can dynamically execute arbitrary code—potentially downloaded or injected during runtime.
        
        -  Security Evasion:  The 
        
        `pipcoloradds-1.0.0.pipcoloradds.\textunderscore \textunderscore init\textunderscore \textunderscore .NoDebugg' node and behavior suggest an attempt to evade debugging or sandbox analysis by detecting or deterring such environments.
        
        -  Cryptography Misuse:  The integration of `fernet.Fernet' may indicate encryption or decryption of stolen data, potentially before exfiltration.
        
        -  Persistence Mechanisms:  The use of `os.makedirs' (`pipcoloradds-1.0.0.pipcoloradds.\textunderscore \textunderscore init\textunderscore \textunderscore .bc\textunderscore initial\textunderscore func.\textunderscore \textunderscore init\textunderscore \textunderscore ') and `shutil.rmtree' 
        
        (pipcoloradds-1.0.0.pipcoloradds.\textunderscore \textunderscore init\textunderscore \textunderscore .bc\textunderscore initial\textunderscore func.bc\textunderscore exit \textunderscore this) suggests the potential setup or cleanup of malicious files or directories for persistence.
        
        Overall, this combination of functionalities strongly aligns with the behavior of malware, such as information stealers or ransomware.",

        \textcolor[RGB]{0,0,0}{``mitigation"}: ``Critical Nodes for Input Validation or Permission Checks:
        
        1. `ctypes.windll.crypt32.CryptUnprotectData' – This is used to decrypt credentials or sensitive data; ensure it cannot access unauthorized data.
        
        2. `httpx.post' and `requests.get'  – These are used for exfiltration; validate and restrict network calls.
        
        3.  `PIL.ImageGrab.grab'  – This is used for screen capture; restrict screen capture permissions.
        
        4.  `pyperclip.copy'  – Prevent unauthorized clipboard manipulation.
        
        5. `exec' – Deny or heavily sandbox dynamic code execution.
        
        6. `fernet.Fernet' – Ensure encryption routines are not used maliciously."
    \}            
\end{tcolorbox}

\begin{tcolorbox}[
    colback=codebg,     
    colframe=black,     
    boxrule=0.5pt,      
    arc=4pt,            
    title= Output Example 3
]
\{\textcolor[RGB]{0,0,0}{``name"}: ``pythonsqliteaddition", \textcolor[RGB]{0,0,0}{``verdict"}: ``Malicious",
        
        \textcolor[RGB]{0,0,0}{``reasoning"}: ``The subgraph structure exhibits multiple indicators of potentially malicious activity:
        
        1. Credential and Data Theft Indicators :
        
        - Calls to `ctypes.windll.crypt32.CryptUnprotectData' are commonly associated with decrypting stored credentials (e.g., browser passwords).
        
        - Functions named `steal\textunderscore cookies', `steal\textunderscore passwords', and `steal\textunderscore screen', combined with calls to `PIL.ImageGrab.grab' (for capturing screenshots) and `re.match' (potentially for parsing sensitive data), strongly imply an intent to steal user information.
        
        - Usage of `pyperclip.copy' within `auto\textunderscore copy\textunderscore wallet.address\textunderscore swap' suggests this action could alter or steal clipboard data, often done in cryptocurrency wallet hijacking.
        
        2. Use of ctypes for Memory Operations:
        
        - The `ctypes.cdll.msvcrt.memcpy' function call could manipulate memory directly, adding another layer of potential obfuscation or tampering.
        
        3. Persistence and Evasion:
        
        - The presence of functions such as `NoDebugg' and `NoDebugg.regkey' indicates anti-debugging and anti-analysis techniques, which are common for malicious software.
        
        - Use of `os.system' and `winreg.QueryValueEx' could be employed for modifying the system environment or registry keys to establish persistence.
        
        4. System Profiling:
        
        - Multiple calls to `psutil' functions such as `psutil.cpu\textunderscore count', `psutil.virtual\textunderscore memory', and `psutil.process\textunderscore iter' suggest the script profiles the victim’s system, which may be used for tailoring attacks or exfiltration.
        
        5. Potential Ransomware Components:
        
        - The use of `bc\textunderscore create\textunderscore file' and `ZipThings', as well as `subprocess.Popen' for zipping or creating files, combined with the functions above, can be indicative of exfiltration or encryption of user data.
        
        6. Exfiltration and Execution:
        
        - URLs are accessed via `urllib.request.Request' and `urllib.request.urlopen', which could be used to send stolen data or download additional malicious components.
        
        - The `finished\textunderscore bc' and `bc\textunderscore exit\textunderscore this' calls ending with `os.\textunderscore exit' may indicate the script’s flow has clear termination after completing malicious operations.",

        \textcolor[RGB]{0,0,0}{``mitigation"}: ``The highest priority nodes or calls for input validation or permission checks are:
        
        1. `ctypes.windll.crypt32.CryptUnprotectData': Limit access to sensitive decryption functions by restricting permissions.
        
        2. `PIL.ImageGrab.grab': Disable unauthorized screenshot functionality.
        
        3. Functions related to clipboard modification (`pyperclip.copy'): Restrict clipboard operations.
        
        4. System commands and registry operations (`os.system', `winreg.QueryValueEx'): Implement strict permission checks and monitor usage.
        
        5. Network-related calls (`urllib.request.Request', `urllib.request.urlopen'): Ensure the script isn’t communicating with malicious entities and validate permitted endpoints.
        
        6. `os.\textunderscore exit': Audit its usage in combination with other suspicious nodes, such as file or data manipulation.
        
        The subgraph strongly indicates malicious intent, and immediate analysis of the full package and its behavior during runtime in an isolated environment is recommended."
    \}            
\end{tcolorbox}

\begin{tcolorbox}[
    colback=codebg,     
    colframe=black,     
    boxrule=0.5pt,      
    arc=4pt,            
    title= Output example 4
]
\{\textcolor[RGB]{0,0,0}{``name"}: ``Flask-Bootstrap", \textcolor[RGB]{0,0,0}{``verdict"}: ``Benign",
        
        \textcolor[RGB]{0,0,0}{``reasoning"}: ``The subgraph provided outlines nodes and their interactions within a Flask-Bootstrap package, which is used to integrate Bootstrap functionality into Flask applications. All the nodes and edges in the subgraph are consistent with expected behavior in this context. The use of WTForms renderer and related DOM elements (`div', `form', `nav', etc.), alongside utilities like `flask.url\textunderscore for', `markupsafe.Markup', and `dominate.tags', indicates standard functionalities for rendering templates, generating URLs, and adding styles or interactivity. There is no evidence of unexpected or suspicious function calls, network interactions, or operations that could potentially mislead or harm the user.",

        \textcolor[RGB]{0,0,0}{``mitigation"}: ``No immediate mitigation is necessary based on this structure. However, as a best practice:  
        
        - Ensure that input validation is implemented on fields rendered by the `WTFormsRenderer' and processed forms to mitigate potential injection attacks.  
        
        - Audit `StaticCDN.get\textunderscore resource\textunderscore url' and `flask.url\textunderscore for' logic to ensure only safe and validated resource URLs are generated or included in templates."
    \}            
\end{tcolorbox}

\begin{tcolorbox}[
    colback=codebg,     
    colframe=black,     
    boxrule=0.5pt,      
    arc=4pt,            
    title= Output example 5
]
\{\textcolor[RGB]{0,0,0}{``name"}: ``CausalInference", \textcolor[RGB]{0,0,0}{``verdict"}: ``Benign",
        
        \textcolor[RGB]{0,0,0}{``reasoning"}: ``The subgraph indicates a package primarily focused on causal inference applications, with utilities and tests leveraging statistical methods and random data generation for validation. Most component nodes and edges point towards expected behavior for a package of this type, such as the use of `numpy.random.multivariate\textunderscore normal' and `numpy.random.binomial' for generating random data, and `distutils.core.setup' for package setup. The connections to various testing nodes and utility functions (e.g., `test\textunderscore causal', `test\textunderscore matching', etc.) suggest an emphasis on proper validation of internal operations. No obvious malicious indicators like obfuscated code executions, network calls, or actions outside the scope of causal inference computations are present.",

        \textcolor[RGB]{0,0,0}{``mitigation"}: ``Although no malicious activity is detected, input validation should be prioritized at the following higher-risk nodes for ensuring package integrity:  
        
        1. `numpy.random.multivariate\textunderscore normal' and `numpy.random.binomial': Validate inputs to random data generation functions to ensure parameters are appropriately bounded to prevent system misuse or instability.  
        
        2. `distutils.core.setup': Confirm that the setup script does not introduce unexpected installation behaviors or dependencies.  
        
        3. `causal.sumlessthan', `calc\textunderscore att\textunderscore se', `log1exp': Ensure robust checks for mathematical operations to prevent improper assertions or computation errors during causal inference calculations."
    \}            
\end{tcolorbox}

\end{document}